\documentclass[12pt]{article}
\usepackage[english]{babel}
\usepackage{amsmath}
\usepackage{cite}
\usepackage{amsfonts}
\usepackage{amssymb}
\begin{document}

\title{
{\bf{}Cubic interaction vertex of higher-spin fields with external
electromagnetic field}}

\author{\sc I.L. Buchbinder, T.V. Snegirev, Yu.M. Zinoviev}
\author{\sc I.L. Buchbinder${}^{ab}$\thanks{joseph@tspu.edu.ru},
T.V. Snegirev${}^b$\thanks{snegirev@tspu.edu.ru}, Yu.M.
Zinoviev$^c$\thanks{Yurii.Zinoviev@ihep.ru}
\\[0.5cm]
\it ${}^a$ The Erwin Schrodinger International Institute for
Mathematical Physics\\
\it A-1090, Vienna, Austria\\[0.3cm]
\it ${}^b$\thanks{Permanent address}{\;}Department of Theoretical
Physics,\\
\it Tomsk State Pedagogical University,\\
\it Tomsk 634061, Russia\\[0.3cm]
\it ${}^c$Institute for High Energy Physics,\\
\it Protvino, Moscow Region, 142280, Russia}
\date{}

\maketitle \thispagestyle{empty}

\begin{abstract}

We fulfill the detailed analysis of coupling the charged bosonic
higher spin fields to external constant electromagnetic field in
first order in external field strength. Cubic interaction vertex of
arbitrary massive and massless bosonic higher spin fields with
external field is found. Construction is based on deformation of
free Lagrangian and free gauge transformations by terms linear in
electromagnetic field strength. In massive case a formulation with
Stueckelberg fields is used. We begin with most general form of
deformations for Lagrangian and gauge transformations, admissible by
Lorentz covariance and gauge invariance and containing some number
of arbitrary coefficients, and require the gauge invariance of the
deformed theory in first order in strength. It yields the equations
for the coefficients which are exactly solved. As a result, the
complete interacting Lagrangian of arbitrary bosonic higher spin
fields with constant electromagnetic field in first order in
electromagnetic strength is obtained. Causality of massive spin-2
and spin-3 fields propagation in the corresponding electromagnetic
background is proved.

\end{abstract}

\newpage

\section{Introduction}
Construction of interacting Lagrangians, which describe coupling of
the higher-spin fields to each other or to low-spin fields or to
external fields is a central line of modern development in
higher-spin field theory. As known, the standard procedure of
switching on the minimal interactions, which are usually used in
low-spin field models, do not work for higher-spin fields. The
attempts of naive adaptations of this procedure to build the
interactions of higher-spin fields yield the inconsistency problems
such as a possibility of propagating non-physical auxiliary fields,
violation of causality, breaking the gauge invariance of free
theory, appearance of ghosts and so on. Some aspects of modern state
of the higher spin field theory are discussed in the reviews
\cite{reviews}.

In general, two types of interaction problems are considered in field
theory, interactions among the dynamical fields and couplings of
dynamical fields to external background. In conventional field
theory, these problems are closely related. However, in higher spin
theory, where the generic interaction Lagrangians are not established
so far, these two types of interactions can be studied as independent
problems (the second one being much simpler).

Consistency problems of higher spin fields couplings to external
electromagnetic field have first been studied in refs. \cite {VZ}.
Inconsistency of higher spin coupling to gravity was investigated in
\cite {AD} for example of spin-2 field. The first attempts to build
the higher spin field Lagrangian interaction have been undertaken in
the refs. \cite{Brink} and \cite{Berends} where the massless higher
spin fields were considered and the consistency aspects were pointed
out. The substantial progress in understanding the massless higher
spin field coupling to gravity has been attained in refs. \cite{FV}
where the it was shown that such coupling demand to involve the
fields of all spins, the higher spin symmetry has been introduced,
the cubic vertex of all spin fields interacting with gravity has
been constructed and it was shown that such a vertex exists in $(A)dS$
space only. Some later, the consistent interacting equations of
motion for all massless higher spin fields have been found \cite{V}.
Recently, the arguments have been given that these equations can be
Lagrangian \cite{HIGHSPINLAGR}.

By the present time, different approaches to constructing the
higher spin field interactions were developed and some progress has
been attained for building the cubic vertex for massless and massive
higher spin fields (see e.g. the recent papers \cite{int1},
\cite{int2}, \cite{int3}, \cite{int4}, \cite{int5} and references
therein).

The aim of this paper is a generic construction of lowest order
interaction vertex for massless and massive bosonic arbitrary spin
fields with constant electromagnetic field. As we pointed out above,
the Lagrangian formulation of higher spin fields in external
background is an independent problem. Just this problem is discussed
in the given paper.

Various aspects of Lagrangian formulation for higher spin fields
in electromagnetic background are discussed in the recent papers
\cite{KZ}, \cite{em}, \cite{PR}, \cite{PRS}. It is worth noting the
related attempts to derive the equations of motion for higher spin
fields in external electromagnetic and gravitational backgrounds from
string theory (see e.g. \cite{AN}, \cite{K1}, \cite{BK}). Generic
problem of constructing the cubic coupling vertex of arbitrary higher
spin fields to electromagnetic field is open.

The paper is organized as follows. Section 2 is devoted to
description of general method to derive the vertices for massless
and massive higher spin fields coupled to constant electromagnetic
field. In Section 3 we solve the problem of cubic coupling of
arbitrary massless integer higher spin field to constant
electromagnetic background and in Section 4 the analogous problem is
solved for massive integer higher spin fields. Conclusion is devoted
to summary of the results obtained. Some technical aspects of
calculations are putted into Appendix.

\section{Procedure of interacting Lagrangian construction for higher
spin fields in electromagnetic background}

In this section we describe a generic scheme for constructing the
cubic interaction vertex of higher spin fields with external
electromagnetic field.

We want to construct the interaction of higher-spin fields with
external constant electromagnetic field strength $F_{\mu\nu}$. As
usual to provide the invariance under the $U(1)_{em}$ group, the e/m
potential $A_\mu$ should enter into a Lagrangian either through
covariant derivative
$$
D_\mu=\partial_\mu-\varepsilon_{U(1)}A_\mu
$$
where $\varepsilon_{U(1)}$ is the $U(1)_{em}$ generator, or through
the electromagnetic field strength $F_{\mu\nu}$
$$
F_{\mu\nu}=\partial_\mu A_\nu-\partial_\nu A_\mu
$$
In the case of external constant electromagnetic field the strength
$F_{\mu\nu}$ is simply a constant antisymmetric matrix and since the
electromagnetic field is external we will not include the kinetic term
for vector field $A_\mu$ into the Lagrangian.

The approach to the vertex construction is based on two points.
First is gauge invariance, the Lagrangian for given model
${\cal{L}}$ is constructed to be invariant with respect to gauge
transformation $\delta$, i.e. the vanishing of variation $\delta
{\cal{L}}=0$\footnote{Up to the total divergence}. Second point is
perturbative consideration where the interacting Lagrangian in
constructed as a sum of terms, which are linear, quadratic and so on
in external field strength. It means that the interacting Lagrangian
is represented as a series in powers of the strength $F$ in the form
$$
{\cal{L}}={\cal{L}}_0+{\cal{L}}_1+...
$$
where ${\cal{L}}_0$ is the free Lagrangian of dynamical fields,
${\cal{L}}_1$ is quadratic in dynamical fields and linear in
strength $F$ and so on. Also, the gauge transformations are written as
the series
$$
\delta=\delta_0+\delta_1+...
$$
where $\delta_0$ are gauge transformations of free theory,
$\delta_1$ are linear in strength $F$ one and so on.

The aim of this paper is to construct in explicit form the first
interacting contribution to the complete Lagrangian ${\cal L}$, i.e.
we consider a first correction to Lagrangian ${\cal{L}}_1$ and first
correction to gauge transformation $\delta_1$. Both these
corrections are linear in strength $F$. The Lagrangian
${\cal{L}}_1$, being quadratic in dynamical fields and linear in
external field, defines the cubic coupling of higher spin fields to
external electromagnetic field. Gauge variation of action in the
case under consideration has the form
\begin{equation}
\label{1} \delta
S=(\delta_0+\delta_1)({\cal{L}}_0+{\cal{L}}_1)=\delta_0
{\cal{L}}_0+\delta_0 {\cal{L}}_1+\delta_1 {\cal{L}}_0+\delta_1
{\cal{L}}_1=0
\end{equation}
Since the variation $\delta_1{\cal L}_1$ is quadratic on $F$, it can
be omitted in first approximation. To find ${\cal L}_1$ in explicit
form we will implement the following procedure. We write down the most
general expressions for gauge transformations ${\delta}_1$ and
Lagrangian ${\cal L}_1$ on the base of Lorentz symmetry and e/m gauge
invariance up to the numerical coefficients. Then the relation
(\ref{1}) yields the equations for the coefficients which can be
solved in principle, although, as we will see bellow, the solutions
are quite non-trivial.

Let us note that there is an important difference between massless and
massive cases. As is known (see e.g. last reference in
\cite{reviews}), for massless fields with spin $s \ge 3/2$ in a flat
Minkowski space it is impossible to switch on minimal e/m interactions
(while it become possible in $(A)dS$ space \cite{em}). At the same
time for massive fields such possibility does exist even in a
Minkowski space with the addition of appropriate non-minimal
corrections with the coefficients proportional to inverse powers of
mass $m$. In this, it is possible to consider a limit when both mass
$m$ and e/m charge $e_0$ simultaneously go to zero in such a way that
only non-minimal terms survive. So our strategy here will be as
follows. In Section 3 we will consider massless case and construct
non-minimal interactions which exist even in Minkowski space. Then, in
Section 4 we will turn to the massive case where as it will be seen
the very same non-minimal interactions play crucial role.

In general, we begin with the free Lagrangian ${\cal{L}}_0$, which
is invariant under free gauge transformation $\delta_0$. Then we add
all admissible corrections to Lagrangian and gauge transformations
${\cal{L}}_1$ and $\delta_1$ respectively and require fulfillment of
the relation (\ref{1}). In the massive case we also include the
minimal interactions.  Essential element of the approach under
consideration is the use of Stueckelberg fields to provide the gauge
invariance in the massive theories (see e.g. \cite{Zi} for metric-like
formalism and \cite{frame} for frame-like one). It is interesting to
point out that the Stueckelberg fields are automatically arose in the
BRST approach to Lagrangian formulation for higher spin fields
\cite{BRST}.

\section{Massless theory}
Before to start a generic analysis we consider the particular cases of
spin-2 and spin-3 fields. It allows us to get some experience of
constructing the interactions of higher spin fields with external
field and apply this experience to coupling of arbitrary higher spin
fields to external electromagnetic fields.

\subsection{Spin-2 field}
We consider the charged massless spin-2 field propagating in
external constant electromagnetic background. Such a field is
described in terms of doublet of rank-2 real symmetric tensor fields
$h_{\mu\nu}{}^i,\;i=1,2$. Free Lagrangian for such theory is well
known and in flat space \footnote{ We work in $d$-dimensional
Minkowski space with metric $g_{\mu\nu}=(+,-,-,-,..)$. The gauge
group is realized as $SO(2)$.} has the form
\begin{equation}\label{2}
{\cal{L}}_{0} = \frac{1}{2} \partial^\alpha h^{\mu\nu,\;i}
\partial_\alpha h_{\mu\nu}{}^i - {(\partial h)}^{\mu,\;i} 
{(\partial h)}_\mu{}^i - (\partial h)^{\mu,\;i} \partial_\mu h^i -
\frac{1}{2} \partial^\mu h^i\partial_\mu h^i
\end{equation}
where $(\partial h)_\mu{}^i=\partial^\nu h_{\mu\nu}{}^i$,
$h^i=g^{\mu\nu}h_{\mu\nu}{}^i$. Lagrangian (\ref{2})
is invariant under standard gauge transformations with gauge
parameter $\xi_\mu{}^i$ 
\begin{equation}\label{3}
\delta_{0} h_{\mu\nu}{}^i = \partial_\mu \xi_\nu{}^i + \partial_\nu
\xi_\mu{}^i
\end{equation}
The equations of motion corresponding to Lagrangian (\ref{2}) are
written as follows
\begin{eqnarray*}
\left(\frac{\delta{\cal{S}}_{0}}{\delta h_{\mu\nu}{}^i}\right) =
-\partial^2 h^{\mu\nu,\;i} + \partial^{\mu} 
(\partial h)^{\nu,\;i} + \partial^{\nu} (\partial h)^{\mu,\;i}
-\partial^\mu\partial^\nu h^i - g^{\mu\nu} (\partial\partial h)^i +
g^{\mu\nu} \partial^2 h^i
\end{eqnarray*}
Condition of gauge invariance of free theory is
$$
\delta_0{\cal{L}}_0 = \delta_{0}h_{\mu\nu}{}^i
\left(\frac{\delta S_0}{\delta h_{\mu\nu}{}^i}\right) = 0
$$
Let us consider the interacting theory in linear approximation in
external field, adding the first order corrections ${\cal{L}}_1$ and
$\delta_1$ to free Lagrangian and free gauge transformation
respectively. Then the condition of gauge invariance (\ref{1}) in
given approximation takes the form
\begin{equation}\label{4}
\delta_{0}h_{\mu\nu}{}^i
\left(\frac{\delta S_1}{\delta h_{\mu\nu}{}^i}\right) + \delta_{1}
h_{\mu\nu}{}^i \left(\frac{\delta S_0}{\delta h_{\mu\nu}{}^i}\right)
= 0
\end{equation}
where $S_1$ is the correction action corresponding to the Lagrangian
${\cal{L}}_1$.

Now, following to scheme, described in Section 2, we should write
down the admissible form of the ${\cal L}_1$ and ${\delta}_1$ up to
numerical coefficients. The most general ansatz for the first
correction to Lagrangian looks like\footnote{As we have already noted,
for the massless fields in Minkowski space it is impossible to switch
on minimal e/m interactions so that electric charge is zero. That is
why throughout this section we use ordinary partial derivatives
instead of covariant ones.}
\begin{equation}\label{5}
{\cal{L}}_1 = \varepsilon^{ij} F^{\alpha\beta} [a_{1} \partial^\mu
h^{\nu,\;i}_\alpha \partial_\mu h_{\nu\beta}{}^j + a_{2}
(\partial h)_\alpha{}^i (\partial h)_\beta{}^j + a_{3} \partial_\alpha
h^{\mu,\;i}_\beta (\partial h)_\mu{}^j + a_{4} (\partial h)_\alpha{}^i
\partial_\beta h^j]
\end{equation}
where $\varepsilon^{ij}=-\varepsilon^{ji},\varepsilon^{12}=1$ and
$a_1, a_2, a_3, a_4$ are the arbitrary real coefficients. The
Lagrangian (\ref{5}) rises the following contribution to equations
of motion
\begin{eqnarray*}
\left(\frac{\delta S_1}{\delta h_{\mu\nu}{}^i}\right) &=&
\varepsilon^{ij} [a_{1} F^{\alpha(\mu} \partial^2 h^{\nu),\;j}_\alpha
+ a_{2} F^{\alpha(\mu} \partial^{\nu)} (\partial h)^j_\alpha
-\frac12 a_{3} F^{\alpha(\mu} \partial_{\alpha}
(\partial h)^{\nu),\;j} + \nonumber\\
&& \quad + \frac12 a_{3} F^{\alpha\beta} \partial_\alpha
\partial^{(\mu} h^{\nu),\;j}_\beta + \frac12 a_{4} F^{\alpha(\mu}
\partial_\alpha \partial^{\nu)} h^j - a_{4} g^{\mu\nu} F^{\alpha\beta}
\partial_\alpha(\partial h)_\alpha^j]
\end{eqnarray*}
where the parentheses denote symmetrization of the indices without
normalization. It is easy to see that the Lagrangian (\ref{5}) is not
invariant under free gauge transformations (\ref{3}). To recover the
gauge invariance we deform the free gauge transformation by the
additional term ${\delta}_1$. The most general ansatz for such
deformation is written as follows
\begin{equation}\label{6}
\delta_{1} h_{\mu\nu}{}^i = \varepsilon^{ij} [\gamma_1 
F_{(\mu}{}^\alpha \partial_\alpha \xi_{\nu)}{}^j + \gamma_2 
F_{(\mu}{}^\alpha \partial_{\nu)} \xi_{\alpha}{}^j + \gamma_3
g_{\mu\nu} F^{\alpha\beta} \partial_\alpha \xi_\beta{}^j]
\end{equation}
where ${\gamma}_1, {\gamma}_2, {\gamma}_3$ are the arbitrary real
coefficients. Recall that in all cases when interacting Lagrangian has
the same or higher number of derivatives as the free one, there always
exists a possibility to make some fields and gauge parameters
redefinitions. In this, all Lagrangians related by such redefinitions
are completely equivalent. In the case at hands we have  a
two-parameter arbitrariness associated with the following
reparametrization of fields and gauge parameters
\begin{eqnarray*}
h_{\mu\nu}{}^i &\Longrightarrow& h_{\mu\nu}{}^i + \kappa_1
\varepsilon^{ij} F_{(\mu}{}^\alpha h_{\nu)\alpha}{}^j \\
\xi_\mu{}^i &\Longrightarrow& \xi_\mu{}^i + \kappa_2 \varepsilon^{ij}
F_{\mu}{}^\alpha \xi_\alpha{}^j
\end{eqnarray*}
Here ${\kappa}_1, {\kappa}_2$ are the arbitrary real coefficients.
This arbitrariness allows us to vanish the terms with coefficients
$\gamma_1$ and $\gamma_2$. Indeed, taking
$\kappa_1=-\gamma_1,\;\kappa_2=\gamma_1-\gamma_2$ we obtain that
first two terms in (\ref{6}) are absent. Calculating the variations
$\delta_0{\cal{L}}_1$, $\delta_1{\cal{L}}_0$ and substituting the
results into (\ref{4}) we obtain the  following relations for the
arbitrary coefficients
\begin{eqnarray*}
&& 2a_1 - a_3 = 0, \qquad 2a_1 + 2a_2 = 0, \qquad 2a_2 - a_3 + 2a_4 =
0 \\
&& a_3 - \gamma_3(d-2) = 0, \qquad a_4 - \gamma_3(d-2) = 0
\end{eqnarray*}
These equations can be easily solved and we find all the
coefficients $a_{1}-a_{4}$ in terms of a single parameter $\gamma_3$
\begin{equation}\label{7}
a_1 = - a_2 = \dfrac12\gamma_3(d-2), \qquad
a_3 = a_4 = \gamma_3(d-2)
\end{equation}

As a result we have constructed the consistent cubic interaction
Lagrangian (\ref{5}) of spin-2 field with electromagnetic field and
the corresponding correction to gauge transformations (\ref{6}). To
be more precise, we obtained the one-parametric family of
Lagrangians and gauge transformations. It is important to note that
the parameter ${\gamma}$ has dimension of inverse mass square.

\subsection{Spin-3 field}

Now let us consider one more example of coupling the higher spin
field to constant external electromagnetic field. It will allow us
to get additional experience in constructing the interaction before
to go to generic case.

The charged spin-3 field is described by doublet of total symmetric
real tensor rank-3 fields $\phi_{\mu\nu\sigma}{}^i,\;i=1,2$. The
free dynamics is described by Lagrangian
$$
{\cal{L}}_0 = -\frac{1}{2} \partial^\alpha 
\phi^{\mu\nu\sigma,\;i} \partial_\alpha\phi_{\mu\nu\sigma}{}^i
+ \frac{3}{2} {(\partial\phi)}^{\mu\nu,\;i} 
{(\partial\phi)}_{\mu\nu}{}^i + 3 {(\partial\partial\phi)}^{\mu,\;i}
\tilde{\phi}_\mu{}^i + \frac{3}{2} \partial^\alpha 
\tilde{\phi}^{\mu,\;i} \partial_\alpha \tilde{\phi}_\mu{}^i
+ \frac{3}{4} (\partial\tilde{\phi})^i (\partial\tilde{\phi})^i
$$
with gauge transformations of the form
$$
\delta_{0} \phi_{\mu\nu\sigma}{}^i = \partial_{(\mu} 
\xi_{\nu\sigma)}{}^i, \qquad \xi^\mu{}_\mu{}^i = 0
$$
here a tilde means a trace of tensor. As in previous subsection, we
begin with writing down the most general first order deformations of
the Lagrangian and gauge transformations.\footnote{In general case
with dynamical e/m field, the higher the spin of particles we want to
consider the higher will be the number of derivatives we will have to
introduce (see e.g third reference in \cite{int2}) so that the cubic
vertex for massless spin $s$ particle contains $(2s-1)$ derivatives.
But in the case of constant e/m field the problem turns out to be less
restricted and as a result has more solutions. In particular, for the
spin 3 case (as well as for arbitrary spin as will be seen later on)
it is enough to consider non-minimal interactions with three
derivatives only.} For the Lagrangian we have
\begin{eqnarray}\label{8}
{\cal{L}}_{1} &=& - \varepsilon^{ij} F^{\alpha\beta} [a_{1}
\partial^\mu \phi_\alpha{}^{\nu\sigma,\;i} \partial_\mu
\phi_{\beta\nu\sigma}{}^j + a_{2} 
(\partial\phi)_\alpha{}^{\nu\sigma,\;i}
(\partial\phi)_{\beta\nu\sigma}{}^j + a_{3} \partial_\alpha
\phi_\beta{}^{\nu\sigma,\;i} (\partial\phi)_{\nu\sigma}{}^j +
\nonumber \\
&& \qquad + a_{4} (\partial\phi)_\alpha{}^{\mu\;i} \partial_\beta
\tilde{\phi}_{\mu}{}^j + a_{5} (\partial\phi)_\alpha{}^{\mu\;i}
\partial_{\mu} \tilde{\phi}_{\beta}{}^j + a_{6} \partial^\mu
\tilde{\phi}_\alpha{}^{i} \partial_\mu \tilde{\phi}_{\beta}{}^j
+ a_{7} \partial_\alpha \tilde{\phi}_\beta{}^{i}
(\partial\tilde{\phi})_{}^j].
\end{eqnarray}
The Lagrangian (\ref{8}) depends on seven real arbitrary numerical
coefficients $a_1, a_2, ..., a_7$. For gauge transformations one can
use, as for the spin-2 case, the arbitrariness in redefinition
of the fields $\phi_{\mu\nu\sigma}{}^i$ and parameters
$\xi_{\mu\nu}{}^i$. It allows us to set
$$
\delta_{1} \phi_{\mu\nu\sigma}{}^i = \gamma \varepsilon^{ij}
g_{(\mu\nu} F^{\alpha\beta} \partial_\alpha \xi_{\sigma)}{}^j
$$
The gauge variation ${\delta}_1$ contains a single arbitrary real
parameter ${\gamma}$. Gauge invariance in linear approximation in
$F$ is written as
\begin{equation}\label{9}
\delta_{0} \phi_{\mu\nu\sigma}{}^i 
\left(\frac{\delta S_1}{\delta \phi_{\mu\nu\sigma}{}^i}\right) +
\delta_{1} \phi_{\mu\nu\sigma}{}^i
\left(\frac{\delta S_0}{\delta \phi_{\mu\nu\sigma}{}^i}\right) = 0
\end{equation}
Equation (\ref{9}) completely defines all unknown coefficients $a_1,
a_2, ..., a_7$ in terms  parameter $\gamma$. The system of equations
for these coefficients follows from (\ref{9}) and has the form
\begin{eqnarray*}
&& 2a_1 - a_3 = 0, \qquad 2a_1 + a_2 = 0, \qquad a_2 - a_3 + a_4 = 0
\\
&& a_2 + a_5 = 0, \qquad 2a_3 - 3\gamma d = 0, \qquad 
a_4 - 3\gamma d = 0 \\
&& a_4 - 2a_7 - \frac{3}{2} \gamma d = 0, \qquad
a_5 - 2a_7 = 0, \qquad a_5 + 2a_6 = 0
\end{eqnarray*}
The solution to this system is
\begin{equation}\label{10}
a_3 = 2a_1 = \dfrac{3}{2}\gamma d, \qquad
a_4 = - 2a_2 = 3\gamma d, \qquad
a_4 = - 2a_6 = 2a_7 = \dfrac{3}{2}\gamma d.
\end{equation}
Arbitrary parameter $\gamma$ has dimension of inverse square mass.

\subsection{Arbitrary integer spin $s$}

Now we generalize the results obtained in previous subsections to
general massless integer spin-$s$ field.

For description of massless charged field of arbitrary integer spin
$s$ we use a doublet of totally symmetric real tensor rank-$s$ fields
$\Phi_{\mu_1\mu_2...\mu_s}{}^i,\;i=1,2$, satisfying also the double
traceless condition
$$
\Phi^{\alpha\beta}{}_{\alpha\beta\mu_1...\mu_{s-4}}{}^i = 0
$$
Further we will use the following compact notations
\begin{equation}\label{11}
\Phi_{\mu_1\mu_2...\mu_s}{}^i = \Phi_s{}^i, \qquad
\partial^{\mu_1} \Phi_{\mu_1s-1}{}^i = (\partial\Phi)_{s-1}{}^i,
\qquad g^{\mu_1\mu_2} \Phi_{\mu_1\mu_2s-2}{}^i = 
\tilde{\Phi}_{s-2}{}^i
\end{equation}
Free theory in these terms is described by Fronsdal Lagrangian
\cite{Fronsdal}
\begin{eqnarray}
\label{12}
{\cal{L}}_0 &=& (-1)^s \frac{1}{2} [\partial^\mu\Phi^{s\;i}
\partial_\mu\Phi_s{}^i - s (\partial\Phi)^{s-1,\;i}
(\partial\Phi)_{s-1}{}^i + s(s-1)  (\partial\Phi)^{\mu_1s-2,\;i}
\partial_{\mu_1} \tilde{\Phi}_{s-2}{}^i - \nonumber\\
&& \qquad - \frac{s(s-1)}{2} \partial^\mu \tilde{\Phi}^{s-2,\;i}
\partial_\mu \tilde{\Phi}_{s-2}{}^i -\frac{s(s-1)(s-2)}{4}
(\partial\tilde{\Phi})^{s-3,\;i} (\partial\tilde{\Phi})_{s-3}{}^i]
\end{eqnarray}
which is invariant under the gauge transformations
\begin{equation}
\label{13}
\delta_{0} \Phi_s{}^i = \partial_{(\mu_1} \xi_{s-1)}{}^i, \qquad
\tilde{\xi}_{s-3}{}^i = 0
\end{equation}
where $\xi_{s-1}{}^i$ is symmetric traceless rank-$(s-1)$ tensor
field and the tilde means a trace.

Further we follow the procedure what was used in two previous
subsections for the cases of spin-2 and spin-3 fields. We write down
the first order corrections ${\cal L}_1$ and ${\delta}_1$ to
Lagrangian and gauge transformation respectively admissible from
Lorentz covariance and e/m gauge invariance. These corrections contain
some number of arbitrary real coefficients. First order gauge
invariance condition has the form
\begin{equation}\label{14}
\delta_{0}\Phi_{s}{}^i
\left(\frac{\delta S_1}{\delta \Phi_{s}{}^i}\right) + \delta_{1}
\Phi_{s}{}^i \left(\frac{\delta S_0}{\delta \Phi_{s}{}^i}\right) = 0
\end{equation}
and imposes the constraints on the coefficients.

We begin with cubic coupling to electromagnetic field. The most
general ansatz for such interacting Lagrangian  has the form
\begin{eqnarray}\label{15}
{\cal{L}}_{1} &=& (-1)^s \varepsilon^{ij} F^{\alpha\beta}
[a_{1} \partial^\mu \Phi_\alpha{}^{s-1,\;i} \partial_\mu
\Phi_{\beta s-1}{}^j + a_{2} (\partial\Phi)_\alpha{}^{s-2,\;i}
(\partial\Phi)_{\beta s-2}{}^j + \nonumber \\
&& \qquad \qquad + a_{3} \partial_\alpha \Phi_\beta{}^{s-1,\;i}
(\partial\Phi)_{s-1}{}^j + a_{4} (\partial\Phi)_\alpha{}^{s-2,\;i}
\partial_\beta \tilde{\Phi}_{s-2}{}^j + \nonumber \\
&& \qquad \qquad + a_{5} (\partial\Phi)_\alpha{}^{\mu_1s-3,\;i}
\partial_{\mu_1} \tilde{\Phi}_{\beta s-3}{}^j + a_{6} \partial^\mu
\tilde{\Phi}_\alpha{}^{s-3,\;i} \partial_\mu 
\tilde{\Phi}_{\beta s-3}{}^j + \nonumber \\
&& \qquad \qquad + a_{7} \partial_\alpha 
\tilde{\Phi}_\beta{}^{s-3,\;i} (\partial\tilde{\Phi})_{s-3}{}^j +
a_{8} (\partial\tilde{\Phi})_\alpha{}^{s-4}
(\partial\tilde{\Phi})_{\beta s-4}{}^j]
\end{eqnarray}
It contains eight terms, where the coefficients $a_1, a_2, ..., a_8$
are the unknown parameters.

Next step is construction of ansatz for correction to gauge
transformations. Recall that, as discussed in above examples, we
have an arbitrariness in redefinition of fields $\Phi_s{}^i$ and
gauge parameters $\xi_{s-1}{}^i$. Using that, one can show that the
correction to gauge transformations can be reduced to
$$
\delta_{1} \Phi_s{}^i = \gamma \varepsilon^{ij} g_{(\mu_1\mu_2}
F^{\alpha\beta} \partial_\alpha \xi_{\beta s-2)}{}^j
$$
Calculating variations $\delta_0{\cal{L}}_{1}$ and
$\delta_1{\cal{L}}_{0}$ and using the gauge invariance relation
(\ref{14}), we obtain the system of algebraic equations for the
unknown coefficients
\begin{eqnarray*}
&& 2a_1 - a_3 = 0, \qquad
2(s-1)a_1 + 2a_2 = 0, \qquad
2a_2 - a_3(s-1) + 2a_4 = 0 \\
&& (s-2)a_2 + a_5 = 0, \qquad
a_3 - \frac{s}{2}\gamma[d+2s-6] = 0 \\
&& a_4 - \frac{s(s-1)}{2}\gamma[d+2s-6] = 0, \qquad
a_4 - 2a_7 - \frac{s(s-1)}{4}\gamma[d+2s-6] = 0 \\
&& a_5 - 2a_7 = 0, \qquad
a_5 + 2a_6 = 0, \qquad
(s-3)a_5 + 4a_8 = 0
\end{eqnarray*}
It is easy to verify that the solution to these equations is
\begin{eqnarray}\label{16}
a_3 &=& 2a_1 = \dfrac{1}{2}\gamma s(d+2s-6) \nonumber \\
a_4 &=& - 2a_2 = \dfrac{1}{2}\gamma s(s-1)(d+2s-6) \nonumber \\
a_4 &=& - 2a_6 = 2a_7 = \dfrac{1}{4}\gamma s(s-1)(s-2)(d+2s-6) \\
a_8 &=& - \dfrac{1}{16}\gamma s(s-1)(s-2)(s-3)(d+2s-6) \nonumber
\end{eqnarray}
The relations (\ref{16}) contains the single arbitrary parameter
$\gamma$ of dimension of inverse mass square. In particular, we see
that for $s=2,3$ the coefficients (\ref{16}) coincides with the ones
for spin-2 and spin-3 fields given in two previous subsections. Note
that the constructed vertex will be useful for finding the cubic
vertex of the massive theory as well.

\section{Massive theory}

{\bfseries Notations.} To simplify the details of calculations we
introduce convenient notations. The Lagrangian and gauge
transformations of the theories under consideration have a common
structure
$$
{\cal{L}} = {\cal{L}}_{00} + {\cal{L}}_{01} + {\cal{L}}_{02} +
{\cal{L}}_{10} + {\cal{L}}_{11} + \dots
$$
$$
\delta = \delta_{00} + \delta_{01} + \delta_{10} + \delta_{11} + \dots
$$
Here the first index in gauge transformations denotes the power of
fields, and one in Lagrangian means the power of fields higher than
quadratic. The second index both in Lagrangian and in
transformations denotes the number of derivatives. That is, in these
general notations we have $\delta_{kn}\sim\partial^n\Phi^k\xi$,
${\cal{L}}_{kn}\sim\partial^n\Phi^{k+2}$, where $\xi$ is gauge
parameter. In such a case the variation of Lagrangian is written as
follows
\begin{eqnarray}\label{17}
\delta
{\cal{L}} &=& \delta_{00} {\cal{L}}_{00} + (\delta_{00}{\cal{L}}_{01}
+ \delta_{01} {\cal{L}}_{00}) + (\delta_{00} {\cal{L}}_{02} +
\delta_{01} {\cal{L}}_{11}) + \delta_{01} {\cal{L}}_{02} + \nonumber
\\
&& + (\delta_{00} {\cal{L}}_{10} + \delta_{10} {\cal{L}}_{00}) +
(\delta_{00} {\cal{L}}_{11} + \delta_{01} {\cal{L}}_{10} +
\delta_{10}{\cal{L}}_{01} + \delta_{11}{\cal{L}}_{00}) + \dots
\end{eqnarray}
Here in the right-hand side the variations are grouped in such a way
that the sum of the first and second indices for each group in
braces has the same value. Therefore the gauge invariance $\delta
{\cal{L}}=0$ requires vanishing for each group of variations
independently that allows us to construct the interaction vertices,
in principle, in any order. We are interested only in cubic
interaction, which corresponds in our notations to the sum of the
first indices of variations to be equal or less than one. For example,
in this notations, the Lagrangian and gauge transformations
constructed in Section 3 for massless theory have the structure
$$
{\cal{L}} = {\cal{L}}_{02} + {\cal{L}}_{13}\qquad
\delta = \delta_{01} + \delta_{12}
$$
Further all considerations are carried out with help of above
notations. As in massless case, we begin with particular case of
spin-2 and spin-3 fields and then will go to general massive case.

\subsection{Spin 2}

{\bfseries Free theory.} For gauge-invariant description of free
massive spin-2 field theory we need a set of fields
$\Phi^a=\{h_{\mu\nu}{},\;b_\mu{},\;\varphi\}$, first of them is
symmetric where $b_\mu$ and $\varphi$ are the auxiliary Stueckelberg
fields. With notations given above, the Lagrangian for such a theory
has the form:
\begin{equation}\label{18}
{\cal{L}}_{0} = {\cal{L}}_{00} + {\cal{L}}_{01} + {\cal{L}}_{02}
\end{equation}
\begin{eqnarray*}
{\cal{L}}_{02} &=& \frac{1}{2} \partial^\alpha h^{\mu\nu}
\partial_\alpha h_{\mu\nu} - {(\partial h)}^\mu {(\partial h)}_\mu +
(\partial h)^\mu \partial_\mu h - \frac{1}{2} \partial^\mu h
\partial_\mu h - \\
&& - \frac{1}{2} \partial^\mu b^{\nu} \partial_\mu b_{\nu} + \frac12
(\partial b) (\partial b) + \frac{1}{2} \partial^\alpha \varphi
\partial_\alpha \varphi \\
{\cal{L}}_{01} &=& m [\alpha_1 h^{\mu\nu} \partial_\mu b_\nu -
\alpha_1 h (\partial b) + \alpha_0 b^\mu \partial_\mu \varphi] \\
{\cal{L}}_{00} &=& m^2 [- \frac{1}{2} h^{\mu\nu} h_{\mu\nu} +
\frac{1}{2} h h + \frac12 \alpha_1 \alpha_0 h \varphi + 
\frac{d}{2(d-2)} \varphi^2] 
\end{eqnarray*}
The Lagrangian (\ref{18}) is invariant under the gauge transformations
$$
\delta_0 = \delta_{00} + \delta_{01}
$$
\begin{eqnarray} 
(\delta_{01} + \delta_{00}) h_{\mu\nu} &=& \partial_{(\mu} \xi_{\nu)}
+ \frac{m\alpha_1}{d-2} g_{\mu\nu} \xi \nonumber \\
(\delta_{01} + \delta_{00}) b_\mu &=& \partial_\mu \xi +
m \alpha_1 \xi_\mu \nonumber \\
\delta_{00} \varphi &=& - m \alpha_0 \xi \nonumber
\end{eqnarray}
where
$$
(\alpha_1)^2 = 2, \qquad (\alpha_0)^2 = 2\frac{d-1}{d-2}
$$
Note that in the massless limit $m\rightarrow0$ this Lagrangian
decomposes into the sum of the Lagrangians describing the massless
fields with spins 2, 1 and 0. Equations of motion corresponding to
Lagrangian (\ref{18}) are written as follows
$$
\left(\frac{\delta{\cal{S}}_{0}}{\delta\Phi^a}\right) =
\left(\frac{\delta{\cal{S}}_{02}}{\delta\Phi^a}\right) +
\left(\frac{\delta{\cal{S}}_{01}}{\delta\Phi^a}\right) +
\left(\frac{\delta{\cal{S}}_{00}}{\delta\Phi^a}\right)
$$
The gauge invariance means vanishing of the variation
$$
\delta_0 {\cal{L}}_{0} = \delta_0 \Phi^a
\left(\frac{\delta{\cal{S}}_{0}} {\delta\Phi^a}\right) = 0
$$

{\bfseries Minimal interaction.} We work in terms of real doublets
$\Phi^a=\{h_{\mu\nu}{}^i,\;b_\mu{}^i,\;\varphi^i\},\;i=1,2$. Let us
introduce a minimal electromagnetic interaction replacing the
ordinary derivatives by covariant ones
\begin{equation}\label{19}
\partial_\mu \rightarrow D_\mu^{ij} = \delta^{ij} \partial_\mu + e_0
\varepsilon^{ij} A_\mu,\qquad \varepsilon^{ij} = -
\varepsilon^{ji},\qquad \varepsilon^{12} = 1,
\end{equation}
Their commutator proportional to e/m field strength
$[D_\mu^{ik},D_\nu^{kj}]=e_0\varepsilon^{ij}F_{\mu\nu}$
($F_{\mu\nu}=\partial_\mu A_\nu-\partial_\nu A_\mu$). The parameter
$e_0$ is a charge. As a consequence, gauge invariance is violated,
and non-invariant part in linear approximation on $F_{\mu\nu}$ is
equal to
\begin{equation}\label{20}
\bar{\delta}_{0} {\bar{\cal{L}}}_{{0}} = ({\delta}_{00}
{\bar{\cal{L}}}_{{02}} + \bar{\delta}_{01} {\bar{\cal{L}}}_{{01}}) +
\bar{\delta}_{01} {\bar{\cal{L}}}_{{02}}
\end{equation}
\begin{eqnarray*}
({\delta}_{00} {\bar{\cal{L}}}_{{02}} + \bar{\delta}_{01}
{\bar{\cal{L}}}_{{01}}) &=& m e_0\varepsilon^{ij} \xi_\mu^i
[ - \alpha_1 F^{\alpha\mu} b_\alpha^j]; \nonumber \\
\bar{\delta}_{01} {\bar{\cal{L}}}_{{02}} &=& e_0 \varepsilon^{ij}
\xi_{\mu}^i [- 4 F^{\alpha\beta} \partial_\alpha 
h_\beta^{\mu,\;j} - 2 F^{\alpha\mu} (\partial h)_\alpha^j + 3
F^{\alpha\mu} \partial_\alpha h^j] + \nonumber \\
&& + e_0 \varepsilon^{ij} \xi^i [2 F^{\alpha\beta} \partial_\alpha
b_\beta^j];
\end{eqnarray*}
where the bar means that we have replaced the ordinary derivatives
by covariant ones.

{\bfseries Non-minimal interaction.} To recover gauge invariance we
add the non-minimal terms to Lagrangian and to gauge
transformations. As we have already mentioned, we assume that the free
theory is deformed in such a way that the massless limit
$e_0\rightarrow0,\;m^2\rightarrow0,\;\frac{e_0}{m^2}=const$ exists.
Therefore it is natural to search the corrections to Lagrangian and
gauge transformations in the form\footnote{Note that in this section
in all terms linear in e/m field strength $F_{\mu\nu}$, i.e. cubic
terms in the Lagrangians and linear terms in gauge transformations, we
will use ordinary partial derivatives because their replacement by
covariant ones produce corrections of order $F^2$ that go beyond the
approximation considered.}
$$
{\cal{L}}_1 = {\cal{L}}_{13} \qquad \delta_1 = \delta_{12}
$$
$$
\delta_{12} h_{\mu\nu}^i =  \frac{\gamma_2}{m^2} \varepsilon^{ij}
g_{\mu\nu} F^{\alpha\beta} \partial_{\alpha} \xi_\beta^j
$$
\begin{eqnarray*}
{\cal{L}}_{13} &=& \frac{e_0}{m^2} \varepsilon^{ij} F^{\alpha\beta}
[a_1 \partial^\mu h^{\nu,\;i}_\alpha \partial_\mu h_{\nu\beta}^j +
a_2 (\partial h)^i_\alpha (\partial h)_\beta^j + a_3 \partial_\alpha
h^{\mu,\;i}_\beta (\partial h)_\mu^j + \nonumber\\
&& \qquad\qquad + a_4 (\partial h)^i_\alpha \partial_\beta h^j + b_1
\partial^\mu b_\alpha^i \partial_\mu b_\beta^j + b_2 \partial_\alpha
b_\beta^i (\partial b)^j]
\end{eqnarray*}
Here $\gamma_2, a_1, a_2, a_3, a_4, b_1, b_2$ are the real
coefficients. Condition of gauge invariance (\ref{1}) looks in this
case like
$$
\bar{\delta}_{0} {\bar{\cal{L}}}_{{0}} + \delta_{0} {\cal{L}}_{1} +
\delta_{1} {\cal{L}}_{0} = 0
$$
or if rewrite it in details
\begin{eqnarray}\label{21}
&& ({\delta}_{00} {\bar{\cal{L}}}_{{02}} + \bar{\delta}_{01}
{\bar{\cal{L}}}_{{01}}) + \bar{\delta}_{01} {\bar{\cal{L}}}_{{02}}
+ (\delta_{00} + \delta_{01}) {\cal{L}}_{13} + \delta_{12}
({\cal{L}}_{00} + {\cal{L}}_{01} + {\cal{L}}_{02}) = \nonumber\\
&=& ({\delta}_{00} {\bar{\cal{L}}}_{{02}} + \bar{\delta}_{01}
{\bar{\cal{L}}}_{{01}}) + (\bar{\delta}_{01} {\bar{\cal{L}}}_{{02}}
+ \delta_{12} {\cal{L}}_{00}) + \nonumber\\
&& + (\delta_{00} {\cal{L}}_{13} + \delta_{12} {\cal{L}}_{01}) +
(\delta_{01} {\cal{L}}_{13} + \delta_{12} {\cal{L}}_{02}) = 0
\end{eqnarray}
Here we have grouped the variations in their index dimensions. Then
to vanish all variations we should vanish each group independently.
First, note that the condition
$$
(\delta_{01} {\cal{L}}_{13} + \delta_{12} {\cal{L}}_{02}) = 0
$$
corresponds exactly to the cubic vertex for the massless theory. Using
the result (\ref{7}) we have
$$
a_1 = - a_2 = \frac12 \gamma_2 (d-2) \qquad a_3 = a_4 = \gamma_2(d-2)
$$
Also it is easy to see that
$$
b_2 = 2b_1
$$
Let us now look at the rest of the relations (\ref{21}). First group
contains only non-invariant part, remaining after the minimal
interaction, and is not compensated, thus we can immediately conclude
that the constructed corrections are not enough. Besides, it is easy
to check that the third group does not vanish, thus it also requires
some new corrections for compensation. This can be done only by
means of contributions from the variations
$\delta_{01}{\cal{L}}_{12}$ and $\delta_{11}{\cal{L}}_{02}$.
Therefore we should build the corrections ${\cal{L}}_{12}$ and
$\delta_{11}$. And, at last, the further calculations show that we
need also the corrections of the form ${\cal{L}}_{11}$ and
$\delta_{11}$. We construct all the corrections in the most general
form. As to the gauge transformations, the only possibility is
$$
\delta_{11} b_\mu^i = \frac{\delta_1}{m} \varepsilon^{ij}
F_{\mu}{}^{\alpha} \xi_\alpha^j
$$ 
Additional terms to the Lagrangian are chosen as follows
\begin{eqnarray}
{\cal{L}}_{12} &=& \frac{1}{m} \varepsilon^{ij} F^{\alpha\beta}
[c_{1} \partial_\alpha h^{\mu,\;i}_\beta b_\mu^j + c_{2} 
(\partial h)^i_\alpha b_\beta^j + c_{3} \partial_\alpha h^i b_\beta^j
+ c_{4} \partial_\alpha b^i_\beta \varphi^j] \nonumber \\
{\cal{L}}_{11} &=& \varepsilon^{ij} F^{\alpha\beta} [d_1 
h^{\mu,\;i}_\alpha h_{\mu\beta}^j + d_2 b^i_\alpha b^j_\beta]
\nonumber
\end{eqnarray}
where $c_{1,2,3,4}, d_{1,2}$ are the arbitrary coefficients.
Further problem is reduced to finding all variations and fulfillment
of three relations
\begin{eqnarray}\label{22}
\delta_{00} {\cal{L}}_{13} + \delta_{01} {\cal{L}}_{12} + \delta_{11}
{\cal{L}}_{02} + \delta_{12} {\cal{L}}_{01} &=& 0 \\
\label{23} {\delta}_{00} {\bar{\cal{L}}}_{{02}} + \bar{\delta}_{01}
{\bar{\cal{L}}}_{{01}} + \delta_{00} {\cal{L}}_{11} + \delta_{11}
{\cal{L}}_{00} &=& 0 \\
\label{24} \bar{\delta}_{01} {\bar{\cal{L}}}_{{02}} + \delta_{00}
{\cal{L}}_{12} + \delta_{01} {\cal{L}}_{11} + \delta_{11}
{\cal{L}}_{01} + \delta_{12}{\cal{L}}_{00} &=& 0
\end{eqnarray}
where we have taken into account the contribution from minimal
interaction (\ref{19}). Relations (\ref{22}) form the system of
algebraic equations, which allows us to express the coefficients
$c_{1,2,3}$ through parameters of gauge transformations
$\gamma_2,\delta_1,b_1$
$$
c_1 = 2 \alpha_1 b_1 - \delta_1 + \alpha_1 \gamma_2(d-1), \qquad
c_2 = - 2 \alpha_1 b_1 - \delta_1, \qquad
c_3 = 2\alpha_1 b_1 + \frac{\delta_1}{2}
$$
The other two relations (\ref{23}), (\ref{24}) yield the following
result
\begin{align*}
& \delta_1 = \frac{2\gamma_2(d-1) - 8 b_1 + 6e_0}{3\alpha_1} &&
d_1 = \frac{-2\gamma_2(d-1) + 2b_1 - 3e_0}{3}, \\
& c_4 = \frac{\alpha_0(\gamma_2(d+2) + 8b_1 - 6e_0}{6}  &&
d_2 = - \frac{e_0}{2}
\end{align*}
Parameters $\gamma_2,\;b_1,e_0$ are arbitrary. As a result the cubic
vertex and the corresponding gauge transformations are found.

{\bfseries Gauge fixing and constraints.} Now we investigate the
causal aspects of constructed Lagrangian, namely we show that the
equations of motion for the field $h_{\mu\nu}{}^i$ contains higher
derivatives only in form of d'Alambertian. First, we fix the gauge
transformations and eliminate the auxiliary fields $b_\mu$ and
$\varphi$. After that the Lagrangian takes the form
\begin{eqnarray*}
{\cal{L}}_{0} &=& \frac{1}{2} D^\alpha h^{\mu\nu} D_\alpha
h_{\mu\nu} - {(D h)}^\mu {(D h)}_\mu - (DD h)h - \frac{1}{2} D^\mu h
D_\mu h -\frac{m^2}{2} h^{\mu\nu} h_{\mu\nu} + \frac{m^2}{2} h h
\nonumber \\
{\cal{L}}_{1} &=& \frac{1}{m^2} \varepsilon^{ij} F^{\alpha\beta}
[a_1 \partial^\mu h^{\nu,\;i}_\alpha \partial_\mu h_{\nu\beta}^j +
a_2 (\partial h)^i_\alpha (\partial h)_\beta^j + a_3 \partial_\alpha
h^{\mu,\;i}_\beta (\partial h)_\mu^j + \nonumber\\
&& \qquad\qquad + a_4 (\partial h)^i_\alpha \partial_\beta  h^j + m^2
d_1 h^{\mu,\;i}_\alpha h_{\mu\beta}^j]
\end{eqnarray*}
It is easy to verify by direct calculations that the equations of
motion yield the usual algebraic constraint $h^i=0$. Indeed, acting by
the following second order differential operator on equations of
motion ones obtain
$$
\left(D^{ik}_\mu D^{kj}_\nu - \frac{m^2} {d-2} \delta^{ij} g_{\mu\nu}
+ \frac{2\delta_1}{m^2 \alpha_1} \varepsilon^{ij} F^\alpha{}_\mu
\partial_\alpha \partial_\nu \right) 
\left(\frac{\delta{\cal{S}}} {\delta h_{\mu\nu}{}^j}\right) = -m^4
\frac{d-1}{d-2} h^i = 0
$$
where we have omitted the terms quadratic in $F$. To see what
happens with the differential constraint in the case under
consideration, we act on the equations of motion by the following
first order differential operator
$$
\left(g_{\mu\alpha} D^{ij}_\nu + \frac{\gamma_2}{2m^2}
\varepsilon^{ij} g_{\mu\nu} F^\sigma{}_\alpha \partial_\sigma \right)
\left(\frac{\delta{\cal{S}}}{\delta h_{\mu\nu}{}^j}\right) = 0.
$$ 
Taking into account that $h^i=0$ one get
$$
- m^2 (Dh)_\alpha{}^i + \varepsilon^{ij} [(2e_0+d_1) F^{\sigma\rho}
\partial_\sigma h_{\alpha\rho}{}^j + (1-d_1) F^{\sigma}{}_\alpha
(\partial h)_{\sigma}{}^j] = 0
$$
We see that the differential constraint is modified and as a result
$(Dh)\sim F,\;(DDh)\sim F^2$. Therefore, up to terms of the second
order in $F$ the equation for $h_{\mu\nu}{}^i$ can be rewritten in the
form
\begin{eqnarray}\label{spin2EoM}
&& - (D^2 h_{\mu\nu})^i - m^2 h_{\mu\nu}{}^i - \dfrac{1}{m^2}
\varepsilon^{ij} F^\alpha{}_{(\mu} (a_1 \partial^2 
h_{\nu)\alpha}{}^j - m^2 d_1 h_{\nu)\alpha}{}^j) + \nonumber \\
&& + \varepsilon^{ij} (2e_0 + d_1 + \dfrac{a_3}{2}) F^{\alpha\beta}
\partial_\alpha \partial_{(\mu} h_{\nu)\beta}{}^j = 0
\end{eqnarray}
If we choose the free parameters so that the coefficient at the last
term is zero, the higher derivatives in this equation form the
d'Alambertian, what guarantees the causality. This can be done by
selecting the appropriate parameter $b_1$
$$
b_1 = \frac{\gamma_2(d+2) - 6e_0}{4}
$$

\subsection{Spin 3}

{\bfseries Free theory.} For gauge invariant description of free
massive spin-3 field theory we need a set of fields
$\Phi^a=\{\phi_{\mu\nu\sigma},h_{\mu\nu}{},b_\mu{},\varphi\}$, first
two of them are totally symmetric. Here $h_{\mu\nu}, b_{\mu}, \varphi$
are the auxiliary Stueckelberg fields. With notations given at
beginning of section 3, the Lagrangian for such a theory has the
form
\begin{equation}\label{25}
{\cal{L}}_{0} = {\cal{L}}_{00} + {\cal{L}}_{01} + {\cal{L}}_{02}
\end{equation}
\begin{eqnarray}
{\cal{L}}_{02} &=& - \frac{1}{2} \partial^\alpha \phi^{\mu\nu\sigma}
\partial_\alpha \phi_{\mu\nu\sigma} + \frac{3}{2}
{(\partial\phi)}^{\mu\nu} {(\partial\phi)}_{\mu\nu} + 3
{(\partial\partial\phi)}^\mu \tilde{\phi}_\mu + \frac{3}{2}
\partial^\alpha \tilde{\phi}^\mu \partial_\alpha \tilde{\phi}_\mu
+ \frac{3}{4} (\partial\tilde{\phi}) (\partial\tilde{\phi}) +
\nonumber \\
&& + \frac{1}{2} \partial^\alpha h^{\mu\nu} \partial_\alpha h_{\mu\nu}
- {(\partial h)}^\mu {(\partial h)}_\mu - (\partial\partial h) h -
\frac{1}{2} \partial^\mu h \partial_\mu h - \nonumber \\
&& - \frac{1}{2} \partial^\mu b^{\nu} \partial_\mu b_{\nu} + \frac12
(\partial b) (\partial b) + \frac{1}{2} \partial^\alpha \varphi
\partial_\alpha \varphi \nonumber \\
{\cal{L}}_{01} &=&  m [\alpha_2 \phi^{\mu\nu\sigma} \partial_\mu
h_{\nu\sigma} - 2 \alpha_2 \tilde{\phi}^\mu {(\partial h)}_\mu
+ \frac{\alpha_2}{2} \tilde{\phi}^\mu \partial_\mu h + \alpha_1
h^{\mu\nu} \partial_\mu b_\nu - \alpha_1 h (\partial b) + \alpha_0
b^\mu \partial_\mu \varphi] \nonumber \\
{\cal{L}}_{00} &=& m^2 [\frac{1}{2} \phi^{\mu\nu\sigma}
\phi_{\mu\nu\sigma} - \frac32 \tilde{\phi}^\mu \tilde{\phi}_\mu -
\frac{\alpha_2\alpha_1}{2} \tilde{\phi}^\mu b_\mu + \frac34 h h +
\frac{\alpha_1\alpha_0}{2} h \varphi - \frac{d+2}{2d}
b^\mu b_\mu + \frac{d+1}{d-2} \varphi^2] \nonumber
\end{eqnarray}
Lagrangian (\ref{25}) is invariant under the gauge transformations
$$
\delta_0 = \delta_{00} + \delta_{01}
$$
\begin{eqnarray*}
(\delta_{01} + \delta_{00}) \phi_{\mu\nu\sigma} &=& \partial_{(\mu}
\xi_{\nu\sigma)} - \frac{2m\alpha_2}{3d} g_{(\mu\nu} \xi_{\sigma)} \\
(\delta_{01} + \delta_{00}) h_{\mu\nu} &=& \partial_{(\mu} \xi_{\nu)}
- m (\alpha_2 \xi_{\mu\nu} - \frac{\alpha_1}{d-2} g_{\mu\nu}\xi) \\
(\delta_{01} + \delta_{00}) b_\mu &=& \partial_\mu \xi + m \alpha_1
\xi_\mu \\
\delta_{00} \varphi &=& - m \alpha_0 \xi
\end{eqnarray*}
where
$$
(\alpha_2)^2 = 3,\qquad (\alpha_1)^2 = 4\frac{d+1}{d}, \qquad
(\alpha_0)^2 = \frac{3d}{d-2}
$$
gauge parameter $\xi_{\mu\nu}$ is symmetric and traceless,
$g^{\mu\nu}\xi_{\mu\nu}=0$. As for the spin-2 case, note that in the
massless limit $m\rightarrow0$ the Lagrangian is decomposed into the
sum of the Lagrangians describing massless fields with spins 3, 2, 1
and 0 that corresponds to a conservation of the number of physical
degrees of freedom.

{\bfseries Minimal interaction.} We formulate the charged spin-3
field in terms of real doublets
$\Phi^a=\{\phi_{\mu\nu\sigma}{}^ih_{\mu\nu}{}^i,\;b_\mu{}^i,\;\varphi^i\},\;i=1,2$. Minimal electromagnetic interaction violates the gauge invariance. Non vanishing gauge variations of the Lagrangian in the linear approximation in $F_{\mu\nu}$ are
\begin{equation}\label{26}
\bar{\delta}_{0} {\bar{\cal{L}}}_{{0}} = ({\delta}_{00}
{\bar{\cal{L}}}_{{02}} + \bar{\delta}_{01} {\bar{\cal{L}}}_{{01}}) +
\bar{\delta}_{01} {\bar{\cal{L}}}_{{02}}
\end{equation}
\begin{eqnarray*}
{\delta}_{00} {\bar{\cal{L}}}_{{02}} +
\bar{\delta}_{01}{\bar{\cal{L}}}_{{01}} &=& m e_0 \varepsilon^{ij}
\xi_{\mu\nu}^i [2 \alpha_2 F^{\alpha\mu} h_\alpha^{\nu,\;j}] +
\nonumber \\
&& +  m e_0 \varepsilon^{ij} \xi_\mu^i [2 \alpha_2 \frac{d+1}{d}
F^{\mu\alpha} \tilde{\phi}_\alpha^j + \alpha_1 F^{\mu\alpha}
b_\alpha^j] \nonumber \\
\bar{\delta}_{01} {\bar{\cal{L}}}_{{02}} &=& 
 e_0 \varepsilon^{ij} \xi_{\mu\nu}^i [6 F^{\alpha\beta} 
\partial_\alpha \phi^{\mu\nu,\;j}_\beta +
6F^{\alpha\mu}(\partial\phi)_{\alpha}^{\nu,\;j} - 6 F^{\alpha\mu}
\partial^\nu \tilde{\phi}_\alpha^j - 9 F^{\alpha\mu} \partial_\alpha
\tilde{\phi}^{\nu,\;j}] + \nonumber \\
&& + e_0 \varepsilon^{ij} \xi_{\mu}^i [-4 F^{\alpha\beta}
\partial_\alpha h_\beta^{\mu,\;j} - 2 F^{\alpha\mu} 
(\partial h)_\alpha^j + 3 F^{\alpha\mu} \partial_\alpha h^j] +
\nonumber \\
&& + e_0 \varepsilon^{ij}\xi^i  [2 F^{\alpha\beta} \partial_\alpha
b_\beta^j]
\end{eqnarray*}
Recall that the bar means that we have replaced the ordinary
derivatives by the covariant ones.

{\bfseries Non-minimal interactions.} To recover gauge invariance we
add non-minimal terms to the Lagrangian and gauge transformations.
Analogously to the spin-2 case, we write down  all possible
additional terms. Corrections to gauge transformations are
$$
\delta_1 = \delta_{11} + \delta_{12}
$$
\begin{align*}
& \delta_{12} \phi_{\mu\nu\sigma}^i = \frac{\gamma_3}{m^2}
\varepsilon^{ij} g_{(\mu\nu} F^{\alpha\beta} \partial_{\alpha}
\xi_{\sigma)\beta}^j && 
\delta_{11} \phi_{\mu\nu\sigma}^i = \frac{\eta_3}{m} \varepsilon^{ij}
g_{(\mu\nu} F_{\sigma)}{}^\alpha \xi_{\alpha}^j \\
& \delta_{12} h_{\mu\nu}^i = \frac{\gamma_2}{m^2} \varepsilon^{ij}
g_{\mu\nu} F^{\alpha\beta} \partial_{\alpha} \xi_\beta^j &&
\delta_{11} h_{\mu\nu}^i =  \frac{\delta_2}{m} \varepsilon^{ij}
F_{(\mu}{}^\alpha \xi_{\nu)\alpha}^j \\
& && \delta_{11} b_\mu^i =  \frac{\delta_1}{m} \varepsilon^{ij}
F_{\mu}{}^\alpha \xi_\alpha^j 
\end{align*}
Additional terms to the Lagrangian look like
\begin{equation}\label{27}
{\cal{L}}_{1} = {\cal{L}}_{11} + {\cal{L}}_{12} + {\cal{L}}_{13}
\end{equation}
\begin{eqnarray*}
{\cal{L}}_{13} &=& - \frac{1}{m^2} \varepsilon^{ij} F^{\alpha\beta}
[a_{11} \partial^\mu \phi^{\nu\sigma,\;i}_\alpha \partial_\mu
\phi_{\nu\sigma\beta}^j + a_{12} (\partial\phi)^{\mu,\;i}_\alpha
(\partial\phi)_{\mu\beta}^j + a_{13} \partial_\alpha
\phi_{\beta}^{\mu\nu,\;i} (\partial\phi)_{\mu\nu}^j + \\
&&\qquad \qquad \quad + a_{14} (\partial\phi)^{\mu,\;i}_\alpha 
\partial_\beta \tilde{\phi}_\mu^j + a_{15} 
(\partial\phi)^{\mu,\;i}_\alpha \partial_\mu \tilde{\phi}_\beta^j +
a_{16} \partial^\mu \tilde{\phi}^i_{\alpha} \partial_\mu
\tilde{\phi}_\beta^j + a_{17} \partial_\alpha \tilde{\phi}^i_\beta
(\partial\tilde{\phi})^j] + \\
&& + \frac{1}{m^2} \varepsilon^{ij} F^{\alpha\beta} [a_{21}
\partial^\mu h^{\nu,\;i}_\alpha \partial_\mu h_{\nu\beta}^j + a_{22}
(\partial h)^i_\alpha (\partial h)_\beta^j + a_{23}\partial_\alpha
h^{\mu,\;i}_\beta (\partial h)_\mu^j + a_{24} (\partial h)^i_\alpha
\partial_\beta h^j] + \\
&& + \frac{1}{m^2} \varepsilon^{ij} F^{\alpha\beta} [b_1\partial^\mu
b_\alpha^i \partial_\mu b_\beta^j + b_2 \partial_\alpha b_\beta^i
(\partial b)^j];  \\
{\cal{L}}_{12} &=& \frac{1}{m} \varepsilon^{ij} F^{\alpha\beta}
[c_{11} \partial_\alpha \phi_\beta^{\mu\nu,\;i} h_{\mu\nu}^j +
c_{12} (\partial\phi)_\alpha^{\mu,\;i} h_{\mu\beta}^j +
c_{13} \partial_\alpha \tilde{\phi}^{\mu,\;i} h_{\mu\beta}^j +
c_{14} \partial_\alpha \tilde{\phi}_\beta^i h^j 
+ c_{15} \partial^\mu \tilde{\phi}_\alpha^i h_{\mu\beta}^j ] + \\
&& + \frac{1}{m} \varepsilon^{ij} F^{\alpha\beta} [c_{21}
\partial_\alpha h^{\mu,\;i}_\beta b_\mu^j + c_{22} 
(\partial h)^i_\alpha b_\beta^j + c_{23} \partial_\alpha h^i
b_\beta^j] +  \\
&& + \frac{1}{m} \varepsilon^{ij} F^{\alpha\beta} [c_{31}
\partial_\alpha b^i_\beta \varphi^j];  \\
{\cal{L}}_{11} &=& \varepsilon^{ij} F^{\alpha\beta} [d_1 
\phi^{\mu\nu,\;i}_\alpha \phi_{\mu\nu\beta}^j + d_2
\tilde{\phi}^i_\alpha \tilde{\phi}_\beta^j + d_3 \tilde{\phi}^i_\alpha
b_\beta^j + d_4 h^{\mu,\;i}_\alpha h_{\mu\beta}^j+ d_5 b^i_\alpha
b^j_\beta];
\end{eqnarray*}
Both the corrections to gauge transformations and corrections to
Lagrangian contain some number of arbitrary parameters.

Before to go further let us make a few comments on above expressions
for Lagrangian.
\begin{itemize}
\item First of all, note that corrections ${\cal{L}}_{13}$ and
$\delta_{12}$ should have the same form as the cubic vertex and gauge
transformations in massless theory. It thus provides the correct
massless limit
$e_0\rightarrow0,\;m^2\rightarrow0,\;\frac{e_0}{m^2}=const$.
\item Corrections to Lagrangian ${\cal{L}}_{12}$ is constructed in
such a way that cross-structures contain the fields describing the
neighboring spins only.
\item To construct the ${\cal{L}}_{11}$ we have considered all
possible combinations of fields. The same concerns the gauge
transformations.
\end{itemize}

Finding the unknown coefficients is based on gauge invariance of the
full Lagrangian ${\cal{L}}={\cal{L}}_{0}+{\cal{L}}_{1}$ with respect
to gauge transformations $\delta=\delta_0+\delta_1$. Also we do not
forget about non-invariant part remaining after the minimal
interaction (\ref{26}). Hence we have
$$
\bar{\delta}_{0} {\bar{\cal{L}}}_{{0}} + \delta_0 {\cal{L}}_{1} +
\delta_1 {\cal{L}}_{0} = 0
$$
or if to rewrite in more details
\begin{eqnarray}\label{paste01}
&& ({\delta}_{00} {\bar{\cal{L}}}_{{02}} + \bar{\delta}_{01}
{\bar{\cal{L}}}_{{01}}) + \bar{\delta}_{01} {\bar{\cal{L}}}_{{02}} +
\nonumber \\
&& + (\delta_{00} + \delta_{01}) ({\cal{L}}_{11} + {\cal{L}}_{12} +
{\cal{L}}_{13}) + (\delta_{11} + \delta_{12})({\cal{L}}_{00} +
{\cal{L}}_{01} + {\cal{L}}_{02}) = \nonumber\\
&=& ({\delta}_{00} {\bar{\cal{L}}}_{{02}} + \bar{\delta}_{01}
{\bar{\cal{L}}}_{{01}} + \delta_{00} {\cal{L}}_{11} + \delta_{11}
{\cal{L}}_{00}) + \nonumber\\
&& + (\bar{\delta}_{01} {\bar{\cal{L}}}_{{02}} + \delta_{00}
{\cal{L}}_{12} + \delta_{01} {\cal{L}}_{11} + \delta_{11}
{\cal{L}}_{01} + \delta_{12} {\cal{L}}_{00}) + \nonumber\\
&& + (\delta_{00} {\cal{L}}_{13} + \delta_{01}{\cal{L}}_{12} +
\delta_{11} {\cal{L}}_{02} + \delta_{12} {\cal{L}}_{01}) +
(\delta_{01}{\cal{L}}_{13} + \delta_{12} {\cal{L}}_{02}) = 0
\end{eqnarray}
Here on the last step, we have grouped the variations by dimensions,
what means in our notations equality the sums of first and second
indices in individual variations. Vanishing the all variations
requires vanishing of each group independently. The rest
consideration is quire direct but rather tedious, therefore we
formulate here only the final results. The intermediate calculations
are given in the Appendix.

Thus, the condition
$$
\delta_{01} {\cal{L}}_{13} + \delta_{12} {\cal{L}}_{02} = 0
$$
corresponds to cubic vertices of massless theory for fields
$\phi_{\mu\nu\sigma}{}^i$ and $h_{\mu\nu}{}^i$. Coefficients in
Lagrangian ${\cal{L}}_{13}$ are given by (\ref{7}) and (\ref{10}).
Following relation 
$$
\delta_{00} {\cal{L}}_{13} + \delta_{01} {\cal{L}}_{12} +
\delta_{11}{\cal{L}}_{02} + \delta_{12} {\cal{L}}_{01} = 0
$$
allow us to express the coefficients in Lagrangian ${\cal{L}}_{12}$
(except $c_{31}$, which in this approximation does not rise the
contribution) through the parameters
$\gamma_3,\gamma_2,\eta_3,\delta_1$ (see Appendix (\ref{33}))
\begin{align*}
& c_{11} = \frac{e_1\gamma_3(d+2) + 3\eta_3d}{2}, &&
c_{21} = 2\alpha_1b_1 + \alpha_1\gamma_2(d-1) - \delta_1, \\
& c_{12} = - \alpha_2\gamma_3d + 3\eta_3d, &&
c_{22} = 2\alpha_1b_1 - \delta_1, \\
& c_{13} = \frac{\alpha_2\gamma_3d + \alpha_2\gamma_2(d-2) - 
3\eta_3d}{2},
&& c_{23} = - 2\alpha_1b_1 + \frac{\delta_1}{2} \\
& c_{14} = \frac{\alpha_2\gamma_3(d-4) - 6\eta_3d}{4}, && {} \\
& c_{15} = \alpha_2\gamma_3d - 3\eta_3d && {}
\end{align*}
Besides, it yields one equation on the parameters of gauge
transformations
\begin{equation}\label{28}
\alpha_2\gamma_3d - \alpha_2\gamma_2(d-2) - 3\eta_3d+2\delta_2 = 0
\end{equation}
From the remaining two relations
$$
{\delta}_{00} {\bar{\cal{L}}}_{{02}} + \bar{\delta}_{01}
{\bar{\cal{L}}}_{{01}} + \delta_{00} {\cal{L}}_{11} +
\delta_{11}{\cal{L}}_{00} = 0
$$
$$
\bar{\delta}_{01} {\bar{\cal{L}}}_{{02}} + \delta_{00} {\cal{L}}_{12}
+ \delta_{01} {\cal{L}}_{11} + \delta_{11} {\cal{L}}_{01} +
\delta_{12} {\cal{L}}_{00} = 0
$$
which lead to the system of equations (\ref{34}), (\ref{35}) (see
Appendix), one can express the other coefficients in terms of
parameters of gauge transformations
\begin{align*}
&
c_{31} = \frac{\alpha_0\gamma_2d}{2} - 
\frac{\alpha_0\delta_1}{\alpha_1}, &&
d_3 = \frac{2\alpha_2\alpha_1b_1 + \alpha_1\delta_2 -
\alpha_2\delta_1}{2}, \\
& d_1 = \frac{3\gamma_3(d+2) + 3\alpha_2\eta_3d + 12e_0}{4},
&& d_4=e_0 \\
& d_2 = \frac{-3\gamma_3d + 3\alpha_2\eta_3d + 2\alpha_2\delta_2 +
6e_0}{4} &&
d_5 = \frac{12b_1(d+2) + 3\alpha_2\eta_3d(d+2) + 2\alpha_2
\delta_2(d+2) - 12de_0}{12d}
\end{align*}
The above two relations impose the conditions on the parameters of
gauge transformations, which together with (\ref{28}) give us the
following solution
\begin{eqnarray*}
&& b_1 = \frac{\gamma_3(d+4) - 2\gamma_2d}{4}, \\
&& \eta_3 = \frac{\alpha_2(\gamma_3(d-2) - \gamma_2(d-2)-6e_0)}{9d},
\\
&& \delta_2 = \frac{-\gamma_3(d+1) + \gamma_2(d-2) - 3e_0}{\alpha_2}
\\
&& \delta_1=\sqrt{\frac{d+1}{d}} \left(\frac{2\gamma_3(d+4) -
2\gamma_2(d+1) + 6e_0}{3}\right)
\end{eqnarray*}
The parameters $\gamma_3$, $\gamma_2$ and $e_0$ are arbitrary. As a
result the first order corrections to Lagrangian and gauge
transformations are found.

{\bfseries Gauge fixing and constraints.} Now we turn to study the
causal aspects of the constructed Lagrangian. The problem is more
complicated in comparison with spin-2 case due to the fact that the
auxiliary scalar field is not eliminated. Indeed, because of
tracelessness of parameter $\xi_{\mu\nu}{}^i$ we cannot kill all
auxiliary fields. For example, one can fix the gauge transformations
so that the  fields $\varphi$, $b_\mu$ are eliminated as well as the
traceless part of tensor $h_{\mu\nu}$, which is determined by the
decomposition
$$
h_{\mu\nu} = h'_{\mu\nu} + \frac{1}{d} g_{\mu\nu} h
$$
Here $h'_{\mu\nu}$ is traceless part, $g^{\mu\nu}h'_{\mu\nu}=0$ and
$h$ is a trace of $h_{\mu\nu}$, $g^{\mu\nu}h_{\mu\nu}=h$. Then,
after such a gauge fixing the Lagrangian takes the form
\begin{eqnarray*}
{\cal{L}}_{0} &=& - \frac{1}{2} D^\alpha \phi^{\mu\nu\sigma} D_\alpha
\phi_{\mu\nu\sigma} + \frac{3}{2} {(D\phi)}^{\mu\nu}
{(D\phi)}_{\mu\nu} + 3 {(DD\phi)}^\mu \tilde{\phi}_\mu + \frac{3}{2}
D^\alpha \tilde{\phi}^\mu D_\alpha \tilde{\phi}_\mu + \\
&& + \frac{3}{4} (D\tilde{\phi}) (D\tilde{\phi}) - 
\frac{(d-2)(d-1)}{2d^2} D^\mu h D_\mu h + \\
&& + m \alpha_2 \frac{d-2}{2d} \tilde{\phi}^\mu D_\mu h +
m^2 [\frac{1}{2} \phi^{\mu\nu\sigma} \phi_{\mu\nu\sigma}
- \frac32 \tilde{\phi}^\mu \tilde{\phi}_\mu + \frac34 h h] 
\end{eqnarray*}
\begin{eqnarray*}
{\cal{L}}_{1} &=& - \frac{1}{m^2} \varepsilon^{ij} F^{\alpha\beta}
(\frac{3\gamma_3d}{4} [\partial^\mu \phi^{\nu\sigma,\;i}_\alpha
\partial_\mu \phi_{\nu\sigma\beta}^j - 2 
(\partial\phi)^{\mu,\;i}_\alpha (\partial\phi)_{\mu\beta}^j +
2 \partial_\alpha \phi_{\beta}^{\mu\nu,\;i} (\partial\phi)_{\mu\nu}^j
+ \\
&&\qquad\qquad + 4 (\partial\phi)^{\mu,\;i}_\alpha \partial_\beta
\tilde{\phi}_\mu^j + 2 (\partial\phi)^{\mu,\;i}_\alpha \partial_\mu
\tilde{\phi}_\beta^j - \partial^\mu \tilde{\phi}^i_{\alpha}
\partial_\mu \tilde{\phi}_\beta^j + \partial_\alpha
\tilde{\phi}^i_\beta (\partial\tilde{\phi})^j] - \\
&&\qquad\qquad  - m \frac{c_{11} + c_{13} + c_{14}d - c_{15}}{d}
\partial_\alpha \tilde{\phi}_\beta^i h^j - m^2 [d_1 
\phi^{\mu\nu,\;i}_\alpha \phi_{\mu\nu\beta}^j + d_2
\tilde{\phi}^i_\alpha \tilde{\phi}_\beta^j]); 
\end{eqnarray*}

First, we show that the scalar field $h^i$ is auxiliary, i.e. the
equations of motion lead to $h^i=0$ as their consequence. To see
that, we act by the following operators on equations of motion for
the fields $\phi_{\mu\nu\sigma}{}^i$, $h^i$ and combine them
\begin{eqnarray}\label{29}
A^{ij}_{\mu\nu\sigma}
\left(\frac{\delta{\cal{S}}}{\delta\phi_{\mu\nu\sigma}{}^j}\right)-
\frac{m}{\alpha_2} ({D^2}^{ij} - 2m^2 \frac{d+1}{d-2} \delta^{ij})
\left(\frac{\delta{\cal{S}}}{\delta h^j}\right) = m^5
\frac{\alpha_2(d+1)}{d-2} h^i = 0
\end{eqnarray}
Here the operator $A^{ij}_{\mu\nu\sigma}$ has the form
\begin{eqnarray*}
A^{ij}_{\mu\nu\sigma} &=& (D^{ik}_\mu D^{kl}_\nu - \frac1d
g_{\mu\nu}{D^2}^{il} - m^2 \frac1d g_{\mu\nu} \delta^{il})
D^{lj}_\sigma + \nonumber\\
&& + \frac{1}{m^2} \varepsilon^{ij} \left(\beta_1 g_{\mu\nu}
F^\alpha{}_\sigma \partial_\alpha \partial^2 +
\beta_2 \partial_\mu \partial_\nu F^\alpha{}_\sigma \partial_\alpha +
m^2 \beta_3 g_{\mu\nu} F^\alpha{}_\sigma \partial_\alpha\right)
\end{eqnarray*}
with the following parameters
\begin{eqnarray*}
&& \beta_1 = \dfrac{\gamma_3}{d},\qquad
\beta_3 = \gamma_3 \dfrac{d+2}{2d} - \gamma_2\dfrac12 \\
&& \beta_2 = - \gamma_3(d+2) + \gamma_2(d-1) - 3e_0 = \alpha_2\delta_2
+ \gamma_3 + \gamma_2 
\end{eqnarray*}
Thus, it follows from (\ref{29}) that $h^i=0$ when $m\neq0$. Further
we set the scalar field to be zero.

One can expect that the algebraic constraint is same as for free
theory, namely $\tilde{\phi}_\mu{}^i = \phi^\alpha{}_{\alpha\mu}{}^i =
0$. We show that this is actually true. To do that, we act on the
equations of motion by the following operator
\begin{eqnarray}\label{30}
B^{ij}_{\rho\mu\nu\sigma}
\left(\frac{\delta{\cal{S}}}{\delta\phi_{\mu\nu\sigma}{}^j}\right)+
(- \frac{m}{\alpha_2} D^{ij}_\rho + \varepsilon^{ij} \frac{1}{m}
\rho_0 F^\alpha{}_\rho \partial_\alpha)
\left(\frac{\delta{\cal{S}}}{\delta h^j}\right) = m^4 \frac{d+1}{d}
\tilde{\phi}_\rho{}^j = 0
\end{eqnarray}
where
\begin{eqnarray*}
B^{ij}_{\rho\mu\nu\sigma} &=& (g_{\mu\rho} D^{ik}_\nu D^{kj}_\sigma -
\frac1d g_{\mu\nu} D^{ik}_\rho D^{kj}_\sigma - \frac{m^2}{d}
g_{\mu\rho} g_{\nu\sigma} \delta^{ij}) + \\
&& + \varepsilon^{ij} \frac{1}{m^2} \left(\rho_1 g_{\mu\rho}
F^\alpha{}_\nu \partial_\alpha \partial_\sigma -  \dfrac1d \rho_1
g_{\mu\nu} F^\alpha{}_\rho \partial_\alpha \partial_\sigma +
\dfrac{\gamma_3}{d} g_{\mu\nu} F^\alpha{}_\sigma \partial_\alpha
\partial_\rho + \right. \\
 && \qquad \qquad \left. + \rho_2 F_{\mu\rho} \partial_\nu
\partial_\sigma - \dfrac{m^2}{2} \gamma_3d g_{\mu\nu} F_{\sigma\rho}
\right) 
\end{eqnarray*}
with the parameters
\begin{eqnarray*}
&& \rho_1 = - \dfrac13 (\gamma_3(d+1) - \gamma_2(d-2) + 3e_0) =
\dfrac{\delta_2}{\alpha_2} \\
&& \rho_2 = - \dfrac{1}{3d}(2d_2-3e_0) =
\dfrac{3\gamma_3 - 3\alpha_2\eta_3 - 2\alpha_2\delta_2d}{6} \\
&& \rho_0 = \dfrac{1}{6\alpha_2} (\gamma_3(5d+2) + \gamma_2(d+4) +
6e_0)= \dfrac{2\gamma_3d + 2\gamma_2 - \alpha_2\eta_3d}{2\alpha_1} 
\end{eqnarray*}
From (\ref{30}) we have the algebraic constraint
$\tilde{\phi}_\rho{}^i = 0$, what will be taken into account later.

Let us study now how the differential constraint is modified. For
that purpose we act on the equations of motion by the first
order differential operator $C^{ij}_{\rho\lambda\mu\nu\sigma}$
$$
C^{ij}_{\rho\lambda\mu\nu\sigma}
\left(\frac{\delta{\cal{S}}}{\delta\phi_{\mu\nu\sigma}{}^j}\right)=0
$$
where
\begin{eqnarray*}
C^{ij}_{\rho\lambda\mu\nu\sigma} &=&
(g_{\mu\rho} g_{\nu\lambda} - \frac1d g_{\rho\lambda} g_{\mu\nu})
D^{ij}_\sigma - \\
&& -\frac{1}{m^2}\varepsilon^{ij}( \frac12 \gamma_3 g_{\mu\nu}
g_{\sigma(\rho} F^\alpha{}_{\lambda)} \partial_\alpha -
\frac{1}{d} \gamma_3 g_{\mu\nu} g_{\rho\lambda} F^\alpha{}_\sigma
\partial_\alpha) 
\end{eqnarray*}
As a result we have the differential constraint
$$
m^2 (D\phi)_{\rho\lambda}{}^i - \varepsilon^{ij} (\frac23 c_1 - 2e_0)
F^{\alpha\beta} \partial_\beta \phi_{\alpha\rho\lambda}{}^j +
\varepsilon^{ij} (\frac23 c_1 + e_0) F^\alpha{}_{(\rho}
(\partial\phi)_{\lambda) \alpha}{}^j = 0 
$$
Taking into account  all constraints, we can  rewrite the equations
of motion (up to the terms quadratic in $F$) in the form
\begin{eqnarray}\label{spin3EoM}
&& D^2 \phi^{\mu\nu\sigma,\;i} + m^2 \phi^{\mu\nu\sigma,\;i}
- \dfrac{1}{m^2} \varepsilon^{ij} [\dfrac{\gamma_3d}{2} \partial^2
F^{\alpha(\mu} \phi^{\nu\sigma),\;j}_\alpha + m^2 \dfrac{2c_1}{3}
F^{\alpha(\mu} \phi_\alpha^{\nu\sigma),\;j}] - \nonumber\\
&& - \dfrac{1}{m^2} \varepsilon^{ij} (\dfrac{\gamma_3d}{2} -
\dfrac23 d_1 + 2e_0) F^{\alpha\beta} \partial_\alpha \partial^{(\mu}
\phi_{\beta}^{\nu\sigma),\;j} = 0
\end{eqnarray}
The last term can be canceled by suitable choice of the free
parameter $\gamma_2$
\begin{equation}\label{gamma}
\gamma_2 = \frac{\gamma_3(d+4) - 6e_0}{d-2}
\end{equation}
 Then we see that the higher derivatives in the equations of motion
(\ref{spin3EoM}) form the d'Alambertian, what ensures the causality.
Note that if we express $\gamma_3$ from (\ref{gamma}) and substitute
in the formula for $b_1$, we get
$$
b_1 = - \frac{\gamma_2(d+2) - 6e_0}{4}
$$ 
that is the same for spin-2 one up to sign redefinition.

\subsection{Arbitrary integer spin}

{\bfseries Free theory.} As before, we use the gauge invariant
approach to massive higher spin theories proposed in \cite{Zi}
(also see first ref. in \cite{KZ}). In this approach the theory of
integer spin $s$ is described by a set of totally symmetric double
traceless tensor fields $\Phi^k$, $\tilde{\tilde{\Phi}}^{k-4}=0$
where $0\leq k\leq s$. We again use the compact notation (\ref{11}).
Free Lagrangian has form
\begin{equation*}
{\cal{L}} = {\cal{L}}_{02} + {\cal{L}}_{01} + {\cal{L}}_{00}
\end{equation*}
\begin{eqnarray}
{\cal{L}}_{02} &=& \sum^s_{k=0}(-1)^k [\frac{1}{2} \partial^\mu
\Phi^k \partial_\mu \Phi_k - \frac{k}{2}
(\partial\Phi)^{k-1}(\partial\Phi)_{k-1} + \frac{k(k-1)}{2}
(\partial\Phi)^{\mu_1k-2} \partial_{\mu_1} \tilde{\Phi}_{k-2} -
\nonumber\\
&& \qquad\qquad - \frac{k(k-1)}{4} \partial^\mu
\tilde{\Phi}^{k-2} \partial_\mu \tilde{\Phi}_{k-2} - 
\frac{k(k-1)(k-2)}{8} (\partial\tilde{\Phi})^{k-3}
(\partial\tilde{\Phi})_{k-3}] \nonumber\\
{\cal{L}}_{01} &=& \sum^s_{k=0} (-1)^k m \left[a_{1k} \Phi^{k-1}
 (\partial\Phi)_{k-1} + a_{2k} \tilde{\Phi}^{k-2} (\partial\Phi)_{k-2}
+ a_{3k} (\partial\tilde{\Phi})^{k-3} \tilde{\Phi}_{k-3}\right]
\nonumber \\
{\cal{L}}_{00} &=& \sum^s_{k=0} (-1)^k m^2 \left[ b_{1k}(\Phi^k)^2
+ b_{2k} (\tilde{\Phi}^{k-2})^2 + b_{3k} \tilde{\Phi}^{k-2} \Phi_{k-2}
\right]\nonumber
\end{eqnarray}
The gauge transformations are written in the form
$$
(\delta_{01} + \delta_{00}) \Phi_k = \partial_{(\mu_1} \xi_{k-1)} +
m \left(\alpha_k \xi_k + \beta_k g_{(\mu_1\mu_2} \xi_{k-2)}\right),
\qquad \tilde{\xi}_{k-3} = 0
$$
All unknown coefficients are uniquely determined by the parameters
$\alpha_k$ from gauge invariance
$$
a_{1k} = - \alpha_{k-1}, \qquad a_{2k} = - \alpha_{k-1}(k-1),
\qquad a_{3k} = - \alpha_{k-1}\frac{(k-1)(k-2)}{4}
$$
$$
b_{1k} = - \frac{1}{2(k+2)}(\alpha_{k+1})^2\frac{d+2k}{d+2k-2}
$$
$$
b_{2k} = \frac{k-1}{4} (\alpha_{k-1})^2 -
\frac{k(k-1)}{8(k+1)}(\alpha_k)^2 \frac{d+2k}{d+2k-4}
$$
$$
b_{3k} = - \frac{1}{2}\alpha_{k-1}\alpha_{k-2},
\qquad \beta_k = \frac{2\alpha_{k-1}}{k(d+2k-6)}
$$ 
Parameter $\alpha_k$ is fixed from the requirement that the
coefficient $b_{1s}$ for basic field ${\Phi}^{s}$ has the standard
canonical value $b_{1s}=-\dfrac12$. Then
$$
(\alpha_{k})^2 = (s-k)(k+1) \frac{(d+s+k-3)}{(d+2k-2)}
$$
Again, note that in the massless limit $m\rightarrow 0$ this
Lagrangian is decomposed in sum of correct Lagrangians describing
the massless fields of all spins below $s$ inclusively.

{\bfseries Minimal interaction.} After switching on the minimal
interactions non-invariant part of variations in linear approximation
on $F^{\mu\nu}$ is equal to
\begin{equation}\label{31}
\bar{\delta} \bar{\cal{L}} = (\delta_{00} + \bar{\delta}_{01})
({\cal{L}}_{00} + \bar{{\cal{L}}}_{01} + \bar{{\cal{L}}}_{02})
\end{equation}
The bar means that we have replaced the usual derivatives by
covariant ones. Here
\begin{eqnarray*}
\delta_{00} \bar{{\cal{L}}}_{02} + \bar{\delta}_{01}
\bar{{\cal{L}}}_{01} &=& \sum_{k=1}^s (-1)^km
e_0 \varepsilon^{ij} \xi_{k-1}{}^i[ -\alpha_{k-1} (k-1)
F^{\alpha\mu_1} \Phi_\alpha{}^{k-2,\;j}] + \\
&& \qquad + (-1)^k m e_0\varepsilon^{ij} \xi_{k-2}{}^i [- \alpha_{k-1}
(k-1)(k-2) \frac{d+2k-5}{d+2k-6} F^{\alpha\mu_1}
\tilde{\Phi}_\alpha{}^{k-3,\;j}] \\
\bar{\delta}_{01} \bar{{\cal{L}}}_{02} &=& \sum_{k=1}^s (-1)^k e_0
\varepsilon^{ij} \xi_{k-1}{}^i [-2k F^{\alpha\beta} \partial_\alpha
\Phi_\beta{}^{k-1,\;j} -
k(k-1)F^{\alpha\mu_1}(\partial\Phi)_\alpha{}^{k-2,\;j} + \\
&& \qquad + \frac{3}{2} k(k-1) F^{\alpha\mu_1} \partial_\alpha
\tilde{\Phi}^{k-2,\;j} + k(k-1)(k-2) F^{\alpha\mu_1} \partial^{\mu_2}
\tilde{\Phi}_\alpha{}^{k-3,\;j}] 
\end{eqnarray*}

{\bfseries Non-minimal interaction.} As in previous subsections to
recover a broken invariance of the Lagrangian we should add the
appropriate corrections to Lagrangian and gauge transformations. The
form and structure of these corrections we have discussed in two
previous subsections for examples of spin-2 and spin-3 fields.
Therefore here we just write down the corresponding deformation
terms for arbitrary spin. The additional Lagrangian has the form
$$
{\cal{L}}_{1} = {\cal{L}}_{13} + {\cal{L}}_{12} + {\cal{L}}_{11}
$$
\begin{eqnarray*}
{\cal{L}}_{13} &=& \sum^s_{k=0} (-1)^k \frac{1}{m^2} \varepsilon^{ij}
F^{\alpha\beta} [c_{1k} \partial^\mu \Phi_\alpha{}^{k-1,\;i}
\partial_\mu \Phi_{\beta k-1}{}^j + c_{2k}
(\partial\Phi)_\alpha{}^{k-2,\;i} (\partial\Phi)_{\beta k-2}{}^j + \\
&& \qquad + c_{3k} \partial_\alpha \Phi_\beta{}^{k-1,\;i}
(\partial\Phi)_{k-1}{}^j + c_{4k} (\partial\Phi)_\alpha{}^{k-2,\;i}
\partial_\beta \tilde{\Phi}_{k-2}{}^j +
c_{5k}(\partial\Phi)_\alpha{}^{\mu_1k-3,\;i} \partial_{\mu_1}
\tilde{\Phi}_{\beta k-3}{}^j + \\
&& \qquad + c_{6k} \partial^\mu \tilde{\Phi}_\alpha{}^{k-3,\;i}
\partial_\mu \tilde{\Phi}_{\beta k-3}{}^j + c_{7k} \partial_\alpha
\tilde{\Phi}_\beta{}^{k-3,\;i} (\partial\tilde{\Phi})_{k-3}{}^j +
c_{8k} (\partial\tilde{\Phi})_\alpha{}^{k-4}
(\partial\tilde{\Phi})_{\beta k-4}{}^j] \\
{\cal{L}}_{12} &=& \sum^s_{k=0}(-1)^k \frac{1}{m} \varepsilon^{ij}
F^{\alpha\beta} [d_{1k} \Phi_\alpha{}^{k-1,\;i} \partial_\beta
\Phi_{k-1}{}^j + d_{2k} \Phi_\alpha{}^{k-2,\;i} 
(\partial\Phi)_{\beta k-2}{}^j + d_{3k} \tilde{\Phi}^{k-2,\;i}
\partial_\alpha \Phi_{\beta k-2}{}^j + \\
&& \qquad + d_{4k} \tilde{\Phi}_\alpha{}^{k-3,\;i}
(\partial\Phi)_{\beta k-3}{}^j + d_{5k} \tilde\Phi_\alpha{}^{k-3,\;i}
\partial_\beta \tilde{\Phi}_{k-3}{}^j + d_{6k}
\tilde\Phi_\alpha{}^{k-4,\;i} (\partial\tilde{\Phi})_{\beta k-4}{}^j]
\\
{\cal{L}}_{11} &=& \sum^s_{k=0} (-1)^k \varepsilon^{ij}
F^{\alpha\beta} \left[ e_{1k} \Phi_\alpha{}^{k-1,\;i}
\Phi_{\beta k-1}{}^j + e_{2k} \tilde{\Phi}_\alpha{}^{k-3,\;i}
\tilde{\Phi}_{\beta k-3}{}^j + e_{3k} \tilde{\Phi}_\alpha{}^{k-3,\;i}
{\Phi}_{\beta k-3}{}^j\right] 
\end{eqnarray*}
All possible linear in external field strength corrections to gauge
transformations (up to possible redefinitions) are written as follows
\begin{eqnarray*}
\delta_{12} \Phi_k{}^i &=& \frac{\gamma_k}{m^2} \varepsilon^{ij}
g_{(\mu_1\mu_2} F^{\alpha\beta} \partial_\alpha \xi_{\beta k-2)}{}^j
\\
\delta_{11} \Phi_k{}^i &=& \frac{1}{m} \varepsilon^{ij} \left(\delta_k
F_{(\mu_1}{}^\alpha \xi_{\alpha k-1)}{}^j + \eta_k g_{(\mu_1\mu_2}
F_{\mu_3}{}^\alpha \xi_{\alpha k-3)}{}^j\right)
\end{eqnarray*}

It remains to vary the above Lagrangian according to rule
(\ref{paste01}) and find unknown coefficients. First of all, note
that the variations with three derivatives
$(\delta_{01}{\cal{L}}_{13}+\delta_{12}{\cal{L}}_{02})$, correspond
to cubic vertex of massless theory. Therefore we immediately can
write down the expressions for the coefficients $c_{1-8,k}$, they
are given by relations (\ref{16}).

Let us consider the variations with two derivatives
$$
\delta_{00} {\cal{L}}_{13} + \delta_{01} {\cal{L}}_{12}
+ \delta_{11} {\cal{L}}_{02} + \delta_{12} {\cal{L}}_{01}
$$ 
They yield the following system of linear equations (\ref{36}) (see
Appendix), allowing to express coefficients $d_{1-6,k}$ through the
parameters of gauge transformations $\gamma$ and $\delta$
\begin{eqnarray*}
&& d_{1k} = \alpha_{k-1}\gamma_k(d+2k-5) - \frac{1}{2} \alpha_{k-1}
\gamma_{k-1} (d+2k-8) - \delta_{k-1} \\
&& d_{2k} = - \frac{1}{2}\alpha_{k-1}\gamma_{k-1}(k-1)(d+2k-8) -
\delta_{k-1}(k-1) \\
&& d_{3k} = - \frac{1}{2}\alpha_{k-1}\gamma_{k-1}(k-1)(d+2k-8)
- \frac{1}{2}(k-1)\delta_{k-1} \\
&& d_{4k} = - \frac{1}{2}\alpha_{k-1}\gamma_{k-1}(k-1)(k-2)(d+2k-8)
- \delta_{k-1}(k-1)(k-2) \\
&& d_{5k} = - \frac{1}{8}\alpha_{k-1}\gamma_k(k-1)(k-2)(d+2k-2) +
\frac14 \alpha_{k-1}\gamma_{k-1}(k-1)(k-2)(d+2k-8) + \\
&&\qquad\quad + \frac{1}{2}\delta_{k-1}(k-1)(k-2) \\
&& d_{6k} = -
\frac{1}{8}\alpha_{k-1}\gamma_{k-1}(k-1)(k-2)(k-3)(d+2k-8) -
\frac{1}{4}\delta_{k-1}(k-1)(k-2)(k-3) 
\end{eqnarray*}
Besides, we have one more equation on parameters what allows to
express $\eta_k$
$$
\eta_k = \frac{\alpha_{k-1} \gamma_{k-1} (d+2k-8) - \alpha_{k-1}
\gamma_k (d+2k-6) + 2 \delta_{k-1}} {k(d+2k-6)}
$$
Let us group the variations with one derivative, and also take into
account the contributions obtained from minimal interaction (\ref{31})
$$
\bar{\delta}_{01} \bar{{\cal{L}}}_{02} + \delta_{00} {\cal{L}}_{12} +
\delta_{01} {\cal{L}}_{11} + \delta_{11} {\cal{L}}_{01} +
\delta_{12} {\cal{L}}_{00}
$$
They yield the system of equations (\ref{37}) (see Appendix), allowing
us to find the coefficients $e_{1-3,k}$ through parameters $\gamma$
and $\delta$
\begin{eqnarray*}
e_{1k} &=& - \frac14 (\alpha_{k-1})^2 \gamma_{k-1} (d+2k-8) +
\frac{k}{k+1} (\alpha_k)^2 \gamma_k \frac{(d+2k-6)(d+2k-3)}{d+2k-4} -
\\
&& - \frac12 \frac{k}{k+1} (\alpha_k)^2 \gamma_{k+1} (d+2k-3)
+ 2 \frac{k}{k+1} \alpha_k \delta_k \frac{d+2k-3}{d+2k-4} -
\alpha_{k-1} \delta_{k-1} + \frac12 ke_0 \\
e_{2k} &=& \frac14 (k-1)(k-2) \left[\frac12 (\alpha_{k-1})^2
\gamma_{k-1} (d+2k-8) - \frac12\frac{k}{k+1}(\alpha_k)^2 \gamma_k
\frac{(d+2k-6)(d+2k)} {d+2k-4} + \right. \\
&& \left. + \frac14 \frac{k}{k+1} (\alpha_k)^2 \gamma_{k+1} (d+2k)
- \frac{k}{k+1} \alpha_k \delta_k \frac{d+2k} {d+2k-4} + 2
\alpha_{k-1} \delta_{k-1} - ke_0\right] \\
e_{3k} &=& - (k-2) [\frac14 \alpha_{k-1} \alpha_{k-2}
\gamma_{k-2}(d+2k-10) + \frac{1}{2} \alpha_{k-1} \delta_{k-2} +
\frac{1}{2} \alpha_{k-2} \delta_{k-1}] 
\end{eqnarray*}

However we did not use yet three equations for the parameters of
gauge transformations (\ref{37}). Taking into account these three
equations for parameters $\gamma_k$ and $\delta_k$  together with
equations which follow from variations without derivatives
(\ref{38}), we obtain finally the recurrent relations for parameters
$\gamma_k$ and $\delta_k$. As a result we have the extremely
complicated system of recurrent relations. Solution to these
relations is written as follows. First of all we introduce the
notations
$$
\hat{\gamma}_s = (d+2s-2)\gamma_s,\qquad \hat{\gamma}_{s-1} = (d+2s-8)
\gamma_{s-1}, \qquad \delta_k = \alpha_k \hat{\delta}_k
$$
Then
\begin{eqnarray*}
\gamma_k &=& \frac{-(s-k-1)\left[3(d+2s-8)(d+2k-2)+4(s-k-3)
(s-k-2)\right]\hat{\gamma}_s} {3(d+2k-2)(d+2k-4)(d+2k-6)} + \\
&& + \frac{(s-k)\left[3(d+2s-6)(d+2k-2)+4(s-k-2)
(s-k-1)\right]\hat{\gamma}_{s-1}} {3(d+2k-2)(d+2k-4)(d+2k-6)} 
\end{eqnarray*}
\begin{eqnarray*}
\hat{\delta}_k &=& \frac{(d+s+k-4)\left[(d+s+k-4)^2+3(s-k-4)
(s-k)+8\right]\hat{\gamma}_s} {3(d+2k)(d+2k-2)(d+2k-4)} - \\
&& - \frac{(d+s+k-3)\left[(d+s+k-3)^2+3(s-k-3)
(s-k+1)+8\right]\hat{\gamma}_{s-1}} {3(d+2k)(d+2k-2)(d+2k-4)} + e_0 
\end{eqnarray*}
The parameters $\gamma_s,\gamma_{s-1}$ are arbitrary. For the cases
$s=2$ and $s=3$ the above relations coincide with expressions for
the parameters given in subsection 2 and subsection 3 respectively.
As a result, the first order corrections to Lagrangian and gauge
transformations for arbitrary spin $s$ are finally found.

\section{Conclusion}\label{summary}
We have constructed the cubic vertex for arbitrary bosonic massless
and massive higher spin fields in $d$ dimensional flat space coupled
to constant electromagnetic background. Construction is based on
gauge-invariant description as a fundamental principle that
determine as basic properties of the free theory as well as possible
forms of interaction. In case of massive theories, to provide the
gauge invariance we have used a formulation with corresponding
Stueckelberg fields.

The cubic interaction vertex is a deformation of free Lagrangian by
the terms linear in electromagnetic field strength. Such a
deformation violates the free gauge invariance and to restore the
gauge invariance of the theory we also must deform the free gauge
transformations by the terms linear in strength. The deformations of
free Lagrangian and free gauge transformation are obtained in
explicit form and gauge invariance of the total Lagrangian has been
tested up to the first order in strength.

The above procedure works in massless theory directly. In the
massive theory, we started with the formulation of the free theory
as the gauge-invariant model. The problem of constructing the
vertex is immediately complicated by the fact that it was necessary
to introduce the auxiliary Stueckelberg fields. Further, the
introduction of the minimal interaction with electromagnetic field
violates gauge invariance. We have constructed the necessary
deformations of the Lagrangian and gauge transformations, so that to
achieve the invariance under the new gauge transformation in first
order in strength.

Both in massless theory and in massive theory the calculations have
been carried out first for spin 2 and spin 3 cases and then the
results were generalized for the fields with arbitrary spins. Also we
have studied the aspects of causality of the interacting theory. We
have proved that the interacting equations of motion for massive spin
2 and spin 3 fields can be identically transformed in such a way that
the higher (second) derivatives form the d'Alambertian. This
circumstance guarantees the causal propagation. Generalization for
arbitrary spin field can be in principle realized in the same way as
for spin 2 and spin 3 fields although is more complicated
technically.

As a result, the general vertex describing cubic coupling of the
arbitrary spin massless and massive fields to constant
electromagnetic field is completely constructed\footnote{Some time
ago Metsaev developed a general classification of cubic vertices in
higher spin theories (see first paper in ref. [9]) using light-cone
formulation. From formal point of view, the vertices constructed in
our paper could be correspond to the case s-s-1 interaction, where
two fields with spin s are both massive or massless and spin-1 field
massless. But similarly to the massless case massive vertex even
restricted to the case of constant e/m field will still contain terms
with up to $(2s-1)$ derivatives so it will not coincide with the
vertex constructed in our paper. As we have already noted the reason
is that for the constant e/m field the problem turns out to be less
restricted and has more solution than the one for dynamical field.
 We grateful to R. Metsaev for discussion of this issue.}.

\section*{Acknowledgments}
I.L.B is thankful to the Erwin Schrodinger Institute, Vienna for
hospitality during his staying at ESI Workshop on "Higher Spin
Gravity", where the paper has been finalized. Also he is
appreciative to M. Porrati, R. Rahman, A. Sagnotti and M. Vasiliev
for valuable discussions and comments. I.L.B and T.V.S are grateful
to the grant for LRSS, project No 224.2012.2 and RFBR grant, project
No 12-02-00121-a for partial support. Work of I.L.B was also
partially supported by RFBR-Ukraine grant, project No 11-02-9045.
Work of Yu.M.Z was supported in parts by RFBR grant No. 11-02-00814.

\appendix
\section*{Appendix}

\section{Calculating of the variations}

During the most part of the paper there was necessary to calculate
the gauge variations of the various first order corrections to free
higher spin Lagrangians. Below we explain and summarize the details
of such calculations.

Let a theory is given in terms of fields $\Phi^a$ and Lagrangian
${\cal L}$ and let the gauge transformations of the fields are
${\delta}{\Phi}^a$. Condition of gauge invariance can be written as
\begin{equation}\label{32}
\delta{\cal{L}} = \delta \Phi^a 
\left(\frac{\delta{\cal{S}}}{\delta\Phi^a}\right) = 0
\end{equation}
Further we present the results of calculating the variations on the
base of (\ref{32}) in the massive theory. These results are
formulated as the systems of equations for the arbitrary parameters
from Lagrangians and gauge transformations.

\subsection{Massive spin-3 field}
Massive charged spin-3 fields is described by a set of fields
$\Phi^a=\{\phi_{\mu\nu\sigma}{}^i,h_{\mu\nu}{}^i,b_\mu{}^i,\varphi^i\}$,
$i=1,2$

Variations with two derivatives give the system of equations
$$
\delta_{00} {\cal{L}}_{13} + \delta_{01} {\cal{L}}_{12} +
\delta_{11}{\cal{L}}_{02} + \delta_{12} {\cal{L}}_{01} = 0
$$
\begin{eqnarray}\label{33}
&& 2\alpha_{2} a_{21} + c_{12} - 2\delta_2 = 0 \nonumber\\
&& 2\alpha_{2} a_{22} + (c_{12} + 2c_{15}) + 2\delta_2 = 0 \nonumber\\
&& \alpha_{2} a_{23} + 2c_{11} + \gamma_3 (2\alpha_2 - 2\alpha_2[d+2])
- 2\delta_2 = 0 \nonumber\\
&& \alpha_{2} a_{23} - (c_{12} + 2c_{13}) = 0 \nonumber\\
&& \alpha_{2} a_{24} - 2c_{14} - \frac{\gamma_3}{2}(2\alpha_2 +
\alpha_2[d+2]) - 2\delta_2 = 0 \nonumber\\
&& \frac{2\alpha_2}{3d}(2a_{11} + a_{15} + 2a_{16}[d+2]) + c_{15} +
3\eta_3d = 0\nonumber\\
&& \frac{2\alpha_2}{3d}(2a_{12} + a_{14}[d+2]) - (2c_{11} - c_{12}) =
0 \nonumber\\
&& \frac{2\alpha_2}{3d}(2a_{12} + a_{15} [d+2]) + c_{12} - 3\eta_3d =
0 \nonumber\\
&& \frac{2\alpha_2}{3d} (a_{13} + a_{14} + a_{17}[d+2]) - c_{13} +
\frac{\gamma_2}{2} (2\alpha_2 - 4\alpha_2 + \alpha_2d) - \frac{3}{2}
\eta_3d = 0\nonumber\\
&& \frac{2\alpha_2}{3d} (a_{13} + a_{14} - 2a_{15} + a_{17}[d+2]) +
(2c_{14} - c_{15}) = 0 \nonumber\\
&& 2\alpha_1 b_1 - c_{22} - \delta_1 = 0 \nonumber\\
&& \alpha_1 b_2 - c_{21} - \gamma_2 (\alpha_1 - \alpha_1d) - \delta_1
= 0 \nonumber\\
&& \alpha_1 b_2 + (c_{22} + 2c_{23}) = 0 \nonumber\\
&& \frac{\alpha_1}{d-2} (2a_{22} + a_{24}d) - (c_{21} - c_{22}) = 0
\end{eqnarray}
Variations with one derivative and contribution from minimal
interaction give the system of equations
$$
\delta_{\bar{0}\bar{1}} {\cal{L}}_{\bar{0}\bar{2}} + (\delta_{00}
{\cal{L}}_{12} + \delta_{01} {\cal{L}}_{11} +
\delta_{12}{\cal{L}}_{00}) = 0
$$
\begin{eqnarray}\label{34}
&& \alpha_2c_{11} - 2d_1 = - 6e_0 \nonumber\\
&& \alpha_2c_{12} + 4d_1 + 2\delta_2\alpha_2 = - 6e_0 \nonumber\\
&& \alpha_2c_{13} + \gamma_3 (3-3[d+2]) - 2\delta_2 \alpha_2 = 9e_0
\nonumber\\
&& \alpha_2c_{15} + 4d_2 - 2\delta_2\alpha_2 = 6e_0 \nonumber\\
&& \alpha_2c_{21} - \frac{\gamma_3}{2} \alpha_2\alpha_1[d+2] -
\delta_2 \alpha_1 = 0 \nonumber\\
&& \alpha_2 c_{22} + 2d_3 - \delta_2 \alpha_1 = 0 \nonumber\\
&& \frac{2\alpha_2}{3d} (2c_{11} + c_{12} + c_{13}[d+2]) -
\alpha_1c_{21} - 2d_4 = 4e_0 \nonumber\\
&& \frac{2\alpha_2}{3d} (c_{11} + c_{14}[d+2]) - \alpha_1c_{23} -
\frac{\eta_3}{2} (2\alpha_2 + \alpha_2[d+2]) + \frac32\gamma_2 d -
\delta_1 \alpha_1 = - 3e_0 \nonumber\\
&& \frac{2\alpha_2}{3d} (c_{12} + c_{15}[d+2]) + \alpha_1 c_{22} +
2\eta_3 (\alpha_2 - \alpha_2[d+2]) - 2d_4 - \delta_1 \alpha_1 = - 2e_0
\nonumber\\
&& \alpha_1c_{31} - \frac{\gamma_2}{2} \alpha_1 \alpha_0 d + \delta_1
\alpha_0 = 0 \nonumber\\
&& \frac{\alpha_1}{d-2} (c_{11} + c_{13} + c_{14}d -c_{15}) + d_3 = 0
\nonumber\\
&& \frac{\alpha_1}{d-2} (c_{21} + c_{22} + c_{23}d) - \alpha_0 c_{31}
+ 2d_5 = e_0
\end{eqnarray}
Variations without derivatives with contribution from minimal
interaction give three equations
$$
\delta_{00} \bar{\cal{L}}_{{0}{2}} + \bar{\delta}_{{0}{1}}
\bar{\cal{L}}_{{0}{1}} + (\delta_{00} {\cal{L}}_{11} +
\delta_{11}{\cal{L}}_{00}) = 0
$$
\begin{eqnarray}\label{35}
&& 2\alpha_{2} d_4 = - 2\sqrt{3} e_0 \nonumber\\
&& \frac{2\alpha_2}{3d} (2d_1 + 2d_2[d+2]) - \alpha_1 d_3 - 3\eta_3
(d+1) + \frac{\delta_1}{2} \alpha_2 \alpha_1 = 2\sqrt{3} \frac{d+1}{d}
e_0 \nonumber\\
&& \frac{2\alpha_2}{3d} d_3[d+2] - 2\alpha_1d_5 +
\frac{\eta_3}{2} \alpha_2 \alpha_1[d+2] + 2\delta_1\frac{d+2}{2d} =
2\sqrt{\frac{d+1}{d}} e_0
\end{eqnarray}

\subsection{Massive field of arbitrary integer spin}
Massive charged field of arbitrary integer spin is described by a
set of fields $\Phi^a=\{\Phi_{k}{}^i\},k=0,...s$, $i=1,2$.

Variations with two derivatives give the system of equations
$$
\delta_{00} {\cal{L}}_{13} + \delta_{01} {\cal{L}}_{12} + \delta_{11}
{\cal{L}}_{02} \delta_{12} {\cal{L}}_{01} = 0
$$
\begin{eqnarray}\label{36}
&& 2\alpha_{k-1} c_{1k-1} + d_{2k} - \delta_{k-1}(k-1) = 0 \nonumber\\
&& \alpha_{k-1} c_{3k-1} + d_{1k}(k-1) + \delta_{k-1}(k-1) +
\gamma_k (-a_{1k}(k-1) + a_{2k}(d+2k-4)) = 0 \nonumber\\
&& \alpha_{k-1} c_{4k-1} - 2d_{5k} + \delta_{k-1}(k-1)(k-2)
+ \gamma_k (a_{1k} \frac{(k-1)(k-2)}{2} - a_{3k} (d+2k-4)) = 0
\nonumber\\
&& 2\alpha_{k-1} c_{2k-1} - (-d_{2k}(k-2) + 2d_{4k}) -
\delta_{k-1}(k-1)(k-2) = 0 \nonumber\\
&& \alpha_{k-1} c_{5k-1} - 2d_{6k} + \delta_{k-1} 
\frac{(k-1)(k-2)(k-3)}{2} = 0 \nonumber\\
&& \alpha_{k-1} c_{3k-1} - (d_{2k} - 2d_{3k}) = 0 \nonumber\\
&& \beta_k (2c_{2k}(k-2) + c_{5k}(d+2k-4)) + d_{2k}(k-2)
+ \eta_k \frac{k(k-1)(k-2)}{2} (d+2k-6) = 0 \nonumber\\
&& \beta_k (c_{1k}(k-1)(k-2) + c_{5k} + 2c_{6k}(d+2k-4)) - d_{4k}
- \eta_k \frac{k(k-1)(k-2)}{2} (d+2k-6) = 0 \nonumber\\
&& \beta_k (c_{3k} \frac{(k-1)(k-2)}{2} + c_{4k}(k-2) -
c_{7k}(d+2k-4)) + d_{3k}(k-2) + \nonumber\\
&&\qquad + \eta_k \frac{k(k-1)(k-2)}{4}(d+2k-6) 
- \gamma_{k-1} (a_{1k} \frac{(k-1)(k-2)}{2} + \nonumber\\
&&\qquad + a_{2k} (k-2) - a_{3k}(d+2k-6)) = 0 \nonumber\\
&& \beta_k (c_{2k}(k-2)(k-3) + 2c_{5k}(k-3) + 2c_{8k}(d+2k-4))
- (d_{4k}(k-3) - 2d_{6k}) - \nonumber\\
&&\qquad - \eta_k\frac{k(k-1)(k-2)(k-3)}{4}(d+2k-6) = 0 \nonumber\\
&& \beta_k(2c_{2k} - c_{4k}(d+2k-4)) - (d_{1k}(k-1) -d_{2k}) = 0
\nonumber\\
&& \beta_k(c_{3k} \frac{(k-1)(k-2)}{2} + c_{4k}(k-2) - 2c_{5k} +
c_{7k}(d+2k-4)) + (d_{4k}+2d_{5k}) = 0
\end{eqnarray}
Variations with one derivatives plus contributions from minimal
interaction give
$$
\delta_{\bar{0}\bar{1}} {\cal{L}}_{\bar{0}\bar{2}} + (\delta_{00}
{\cal{L}}_{12} + \delta_{01} {\cal{L}}_{11} + \delta_{12}
{\cal{L}}_{00}) = 0
$$
Using notations
$$
e_1 = (k-1)e_0, \qquad e_2 = (k-1)(k-2)e_0, \qquad 
e_3 = (k-1)(k-2)(k-3)e_0
$$
we obtain:
\begin{eqnarray}\label{37}
&& \alpha_{k-1} d_{1k-1} - \delta_{k-1} a_{1k-1} + \gamma_k
b_{3k}(d+2k-4) = 0 \nonumber\\
&& \alpha_{k-1} d_{2k-1} - 2e_{3k} + \delta_{k-1} a_{1k-1}(k-2) = 0
\nonumber\\
&& \alpha_{k-2} d_{1k-1} - \beta_k (d_{1k}(k-1) + d_{2k} -
d_{3k}(d+2k-4)) + 2e_{1k-1} = - 2e_1 \nonumber\\
&& \alpha_{k-2} d_{2k-1} + \beta_k (d_{2k}(k-2) - d_{4k}(d+2k-4)) -
2e_{1k-1} (k-2) + \nonumber\\
&&\qquad + \delta_{k-2} a_{1k-1} (k-2) + \eta_k (a_{1k}(k-1)(k-2) - 
a_{2k} (k-2) (d+2k-4)) = - e_2 \nonumber\\
&& \alpha_{k-2} d_{3k-1} + \beta_k (d_{1k} \frac{(k-1)(k-2)}{2} -
d_{5k}(d+2k-4)) + \delta_{k-2} a_{2k-1} - \nonumber\\
&&\qquad - \eta_k (a_{1k} \frac{(k-1)(k-2)}{2} - a_{3k}(d+2k-4)) +
\nonumber\\
&&\qquad + \gamma_{k-1} (b_{1k-1} (k-1)(k-2) - 2b_{2k-1} (d+2k-6)) =
-\frac{3}{2} e_2 \nonumber\\
&& \alpha_{k-2} d_{4k-1} - \beta_k (d_{2k} \frac{(k-2)(k-3)}{2} -
d_{6k} (d+2k-4)) + 4e_{2k-1} + \nonumber\\
&&\qquad + \delta_{k-2} a_{2k-1}(k-3) - \eta_k (a_{1k}
\frac{(k-1)(k-2)(k-3)}{2} - \nonumber\\
&&\qquad - a_{3k}(k-3)(d+2k-4)) = - e_3 \nonumber\\
&& \beta_{k-1} (d_{1k} \frac{(k-1)(k-2)}{2} + d_{3k} (k-2) - d_{4k} +
d_{5k}(d+2k-6)) - e_{3k} = 0 \nonumber\\
&& \beta_{k-1} (d_{2k} \frac{(k-2)(k-3)}{2} + d_{4k} (k-3) +
d_{6k}(d+2k-6)) - e_{3k}(k-3) + \nonumber\\
&&\qquad + \eta_{k-1} (a_{1k} \frac{(k-1)(k-2)(k-3)}{2} + a_{2k}
(k-2)(k-3) - \nonumber\\
&&\qquad - a_{3k} (k-3)(d+2k-6)) = 0
\end{eqnarray}
Variations without derivatives with contribution from minimal
interaction give three equations
$$
\delta_{00} \bar{\cal{L}}_{{0}{2}} + \bar{\delta}_{{0}{1}}
\bar{\cal{L}}_{{0}{1}} + (\delta_{00} {\cal{L}}_{11} +
\delta_{11}{\cal{L}}_{00}) = 0
$$
\begin{eqnarray}\label{38}
&& - 2\alpha_{k-2} e_{1k-2} - \beta_k e_{3k} (d+2k-4) -2 \delta_{k-2}
b_{1k-2} (k-2) - \eta_k b_{3k} (k-2) (d+2k-4) = \nonumber\\
&& = - \alpha_{k-2} (k-2) e_0 \nonumber\\
&& - \alpha_{k-2} e_{3k} - \beta_k (e_{1k}(k-1)(k-2) +
2e_{2k}(d+2k-4)) - \delta_{k-2} b_{3k}(k-2) - \nonumber\\
&& - \eta_k(b_{1k} k(k-1)(k-2) + 2b_{2k} (k-2) (d+2k-4)) = 
\alpha_{k-1} \frac{d+2k-5} {d+2k-6} e_2
\end{eqnarray}

\end{document}